\documentclass[aps,prl,twocolumn,showpacs,superscriptaddress]{revtex4-2}
\usepackage{latexsym}
\usepackage{amssymb}
\usepackage{graphicx}
\usepackage{amsmath}
\usepackage{bm}
\usepackage[colorlinks,
          linkcolor=black,
            citecolor=black,
            urlcolor=blue
           ]{hyperref}
\usepackage{verbatim}
\usepackage{mathrsfs}
\usepackage{extarrows}
\usepackage{comment}
\usepackage{mathtools,slashed}

\usepackage{cancel}
\usepackage{subfigure}
\usepackage{graphicx}
\usepackage{caption}
\usepackage{ragged2e}

\begin{document}

\title{Local Thermodynamic Reconstruction of Impurity Entropy}

\author{Hang Su}
\email{suhang980715@gmail.com}
\affiliation{Physics Department, King Fahd University of Petroleum and Minerals (KFUPM), Dhahran, Saudi Arabia}
\affiliation{Quantum Center, KFUPM, Dhahran, Saudi Arabia}

\begin{abstract}
Resolving the thermodynamic entropy associated with a local impurity in a spatially extended interacting quantum bath remains a numerical challenge.
We develop a Local-Maxwell approach that reconstructs the impurity entropy from local observables in worldline quantum Monte Carlo (QMC) simulations. 
By integrating their temperature responses along the impurity coupling, the method avoids separate estimates of extensive entropies.
For nonnegative worldline expansions with bounded local couplings, we establish bath-size independent bounds on all factorial moments of the impurity vertex count.
For even vertex counts, the Maxwell estimator variance remains bounded as the coupling approaches zero.
The transverse-field Ising benchmark reproduces exact free fermion results, 
while an interacting XXZ chain demonstrates up to several hundredfold improvement in statistical efficiency over conventional extensive thermodynamic integration. 
In a three-channel XXZ impurity junction, we reveal entropy recovery upon cooling and a nonmonotonic coupling dependence of the finite-temperature entropy contrast.
These results establish local thermodynamic reconstruction as an efficient approach to impurity entropy in interacting quantum baths.
\end{abstract}

\date{\today}


\maketitle

\paragraph{Introduction.---}
Thermodynamic entropy provides a sensitive probe of the degrees of freedom governing the low-temperature behavior of quantum many-body systems~\cite{RevModPhys.80.395}. 
In quantum impurity systems, the impurity entropy $S_{\mathrm{imp}}=S_{\mathrm{full}}-S_{\mathrm{bath}}$ 
characterizes the excess thermodynamic entropy associated with a localized impurity~\cite{PhysRevLett.67.161,Affleck:1992prb,PhysRevLett.93.030402}.
It reveals impurity screening and boundary crossovers, including fractional residual entropies in multichannel Kondo systems~\cite{AffleckLudwig1993,PhysRevLett.128.146803,Gaines:2026aa} and entropy signatures of Majorana zero modes~\cite{PhysRevLett.123.147702}. 
These entropy signatures are of particular interest in mesoscopic quantum devices and engineered quantum impurity systems, 
where quantum dots and charge-Kondo circuits provide tunable platforms for probing strongly correlated and potentially non-Abelian states~\cite{RevModPhys.80.1083,hartman2018direct,Child2022,Karki2023,Piquard2026}.

Despite its physical significance, accurately computing impurity entropy in systems with spatially extended interacting baths remains challenging. 
Numerical renormalization group~\cite{RevModPhys.47.773,RevModPhys.80.395} efficiently captures impurity thermodynamics but is less readily applicable to interacting baths, 
while tensor network methods~\cite{White:1992aa,White:1993aa,Schollwock:2011} can treat such systems explicitly, although finite-temperature calculations can be computationally demanding~\cite{PhysRevB.72.220401}. 
Quantum Monte Carlo (QMC) methods~\cite{prokof1998exact,kawashima:2004aa,Gubernatis:2020,sandvik2010computational} offer a complementary approach to finite-temperature properties of quantum many-body systems, 
but thermodynamic entropy is not generally a directly sampled observable~\cite{PhysRevLett.115.077202}. 
Existing approaches reconstruct entropy through energy based thermodynamic integration (TI), 
including impurity specific QMC implementations~\cite{Huscroft2000,Zitko2012}, 
or partition function based methods  
such as expanded ensembles~\cite{lyubartsev1992new}, reweighting schemes~\cite{Yan:2024} 
and partition function ratios for boundary entropy~\cite{Tang2017} or temperature-incremental calculations~\cite{Dai:2025}.
However, for impurity systems, directly subtracting separately estimated full system 
and bath entropies requires resolving an $\mathcal{O}(1)$ difference between two \textit{extensive} $\mathcal{O}(L)$ quantities. 
Statistical fluctuations in these estimates can consequently obscure the much smaller impurity signal.

An alternative is to exploit thermodynamic Maxwell relations, 
which connect the entropy response to a tunable parameter with the temperature dependence of its conjugate observable~\cite{CooperStern2009}. 
This approach has enabled entropy measurements in mesoscopic quantum dots~\cite{hartman2018direct,Child2022} 
and inspired proposals to probe fractional entropy in multichannel Kondo devices~\cite{PhysRevLett.123.147702,PhysRevLett.128.146803} and entropy signatures of topological phases~\cite{Sankar2023}. 
More recently, entropy reconstruction based on Maxwell relations~\cite{huang2008statistical,callen1960thermodynamics} 
has also been demonstrated numerically in quantum gases using local pair correlations~\cite{PhysRevLett.133.100403} 
and in interacting Hubbard models through thermodynamic integration in parameter space~\cite{Song2026}.  
For impurity systems with explicitly interacting baths, this suggests reconstructing entropy by varying a coupling of bounded spatial support.
A central question is whether the resulting QMC estimator can avoid extensive subtraction while maintaining controlled statistical fluctuations as the bath size increases.

In this Letter, we develop a Local-Maxwell method for reconstructing impurity entropy from local coupling responses in worldline QMC~\cite{prokof1998exact,kawashima:2004aa,Gubernatis:2020,sandvik2010computational,sadoune:2022,Zhang:2024aa,Su:2025}.
The reconstruction uses local impurity vertex counts and their generalization to diagonal coupling terms, 
avoiding the subtraction of independently estimated extensive entropies. 
%
For nonnegative worldline expansions with bounded local couplings, we establish bath-size independent bounds on the factorial moments of the impurity vertex count.
An additional variance bound applies to the Maxwell estimator for even vertex counts. 
We validate the reconstruction against exact results in the critical transverse-field Ising model (TFIM) 
and demonstrate improved statistical efficiency over extensive thermodynamic integration in an interacting XXZ spin chain. 
Applied to a three-channel XXZ impurity junction, the method reveals entropy recovery upon cooling, 
further supported by free energy chord analysis of the finite-temperature response.

\paragraph{Local-Maxwell reconstruction and statistical locality.---}
We start from a Hamiltonian $\mathcal H(\lambda)=\mathcal H_0+\lambda V_D$, 
where $V_D$ is a \textit{local} impurity coupling operator with an operator norm bounded independently of the bath size. 
In a sign problem free worldline QMC expansion, we first consider the case 
where the local impurity exchange enters through vertices carrying a factor of $\lambda$. 
The partition function takes the form~\cite{kawashima:2004aa}
\begin{equation}
\mathcal Z(\beta,\lambda)=\sum_{\mathscr C}W(\mathscr C,\lambda,\beta)=\sum_{\mathscr C}\lambda^{N_D(\mathscr C)}\bar W(\mathscr C,\beta),
\label{eq:local_partition}
\end{equation}
where $\mathscr C$ denotes a worldline configuration, $\beta$ is the inverse temperature (with $k_B=1$), $N_D$ counts the local exchange vertices, 
and the rest $\lambda$-independent weight $\bar W(\mathscr C,\beta)$ is nonnegative. 
The factor $\lambda^{N_D}$ gives $\lambda\partial_\lambda\ln\mathcal Z=\langle N_D\rangle$.
Combining this with the thermodynamic relation $\partial_\lambda\ln\mathcal Z=-\beta\langle V_D\rangle$~\cite{sandvik2010computational}, 
we obtain $\lambda\langle V_D\rangle=-\langle N_D\rangle/\beta$.
The local exchange energy is therefore obtained directly from the impurity vertex count.

By introducing the Maxwell relation $\partial_\lambda S=-\partial_T\langle V_D\rangle$~\cite{huang2008statistical,callen1960thermodynamics}, we obtain
\begin{equation}
\lambda\frac{\partial S}{\partial\lambda}=\langle N_D\rangle-\beta\frac{\partial\langle N_D\rangle}{\partial\beta}.
\label{eq:local_maxwell}
\end{equation}
We thus reconstruct the impurity entropy by integrating this local response,
\begin{equation}
S_{\mathrm{imp}}(\beta,\lambda)=S_{\mathrm{imp}}(\beta,0)+\int_0^\lambda\frac{\mathrm d\lambda'}{\lambda'}\left[\langle N_D\rangle_{\lambda'}-\beta \frac{\partial \langle N_D \rangle_{\lambda'}}{\partial\beta} \right],
\label{eq:entropy_reconstruction}
\end{equation}
where the subscript $\lambda'$ denotes an expectation value evaluated at impurity coupling $\lambda'$. 
For a decoupled spin-$1/2$ impurity, $S_{\mathrm{imp}}(\beta,0)=\ln2$, reflecting its twofold spin degeneracy. 
Since the reference bath is independent of $\lambda$, i.e., $\partial_\lambda S_{\mathrm{imp}}=\partial_\lambda S$, 
the thermal response of the local exchange energy determines the impurity entropy change, including the response of the surrounding bath.
This avoids subtracting two extensive entropy estimates. 

We examine how the equilibrium fluctuations of $N_D$ depend on the bath size.
For the expansion in Eq.~\eqref{eq:local_partition}, the $k$-th factorial moment is given by $\langle(N_D)_k\rangle=\lambda^k\partial_\lambda^k\mathcal Z/\mathcal Z$, 
where $(N_D)_k=N_D(N_D-1)\cdots(N_D-k+1)$ for integer $k\geq1$.
In the operator representation, the same derivatives involve $k$ imaginary-time insertions of $V_D$. 
These insertions are bounded by the operator norm of $V_D$, giving $|\partial_\lambda^k\mathcal Z|\leq\beta^k\mathcal Z\|V_D\|^k$~\cite{bhatia2009positive,bhatia2013matrix,schatten2013norm}, as proved in the End Matter.
For nonnegative worldline weights and $\lambda\geq0$, we obtain $0\leq\langle(N_D)_k\rangle\leq(\beta\lambda\|V_D\|)^k$.
Using $N_D^2=(N_D)_2+N_D$, the first two factorial moments give
\begin{equation}
\operatorname{Var}(N_D)\leq b+b^2,\qquad b=\beta\lambda\|V_D\|,
\label{eq:variance_bound}
\end{equation}
%

For impurity-exchange vertices that flip the impurity spin, with no other spin-flip terms acting on the impurity, worldline periodicity requires $N_D$ to be even.
Since $N_D^2\leq2(N_D)_2$, we obtain the stronger bound
\begin{equation}
\operatorname{Var}\left(\frac{N_D}{\lambda}\right)\leq2\beta^2\|V_D\|^2,\qquad\lambda>0.
\label{eq:variance_bound}
\end{equation}
Both bounds are independent of the bath size at fixed $\beta$ and $\lambda$, while the latter remains uniform as $\lambda\to0^+$.
The proof requires neither a spectral gap nor a finite correlation length and remains valid for critical baths. 
These bounds concern equilibrium fluctuations of local observables, rather than the statistical uncertainty of the reconstructed entropy.
Temperature differentiation can amplify statistical noise, while coupling integration introduces error propagation and discretization effects.

The reconstruction in Eq.~\eqref{eq:entropy_reconstruction} assumes that $\lambda$ enters only through the exchange vertices. 
The coupling may also generally appear in diagonal configuration weights, whose contributions are not captured by $N_D$. 
We define a local estimator $Q_\lambda(\mathscr C)=\partial_\lambda\ln W(\mathscr C,\lambda,\beta)$, 
which can be evaluated directly from each sampled configuration using the local vertex count and the imaginary-time integral of diagonal interactions.
Its expectation value satisfies $\langle Q_\lambda\rangle=\partial_\lambda\ln\mathcal Z$.
The general Local-Maxwell relation becomes
\begin{equation}
\frac{\partial S_{\mathrm{imp}}}{\partial\lambda}=\langle Q_\lambda\rangle-\beta\frac{\partial\langle Q_\lambda\rangle}{\partial\beta}.
\label{eq:general_local_maxwell}
\end{equation}
The temperature derivative acts on the full expectation value, including any explicit $\beta$ dependence of $Q_\lambda$.
For the vertex-only case $W(\mathscr C,\lambda,\beta)=\lambda^{N_D(\mathscr C)}\bar W(\mathscr C,\beta)$, it recovers Eq.~\eqref{eq:local_maxwell} with $Q_\lambda=N_D/\lambda$.

\paragraph{Transverse-field Ising benchmark.---}
We first benchmark the Local-Maxwell method against an exactly solvable TFIM with a boundary impurity~\cite{PhysRevB.60.4195,PhysRevB.96.155145}. 
The Hamiltonian is
\begin{equation} 
	\mathcal{H}_{\text{TFIM}} = -J \sum_{j=1}^{L-1} \sigma^z_j \sigma^z_{j+1} - \Gamma \sum_{j=1}^{L} \sigma^x_j -\lambda \sigma^z_0 \sigma^z_{1},
	\label{eq:Htfim}
\end{equation}
where site $0$ denotes the impurity spin without a transverse field, and $L$ is the number of bath spins. 
Here, we set $\Gamma = J = 1$, placing the bath at its quantum critical point.
The impurity entropy can also be calculated exactly using a free fermion representation~\cite{PhysRev.127.1508,PFEUTY197079,PhysRevB.97.245127}.
\begin{figure}[tbp]
\centering
\subfigure{\includegraphics[width=0.2385\textwidth]{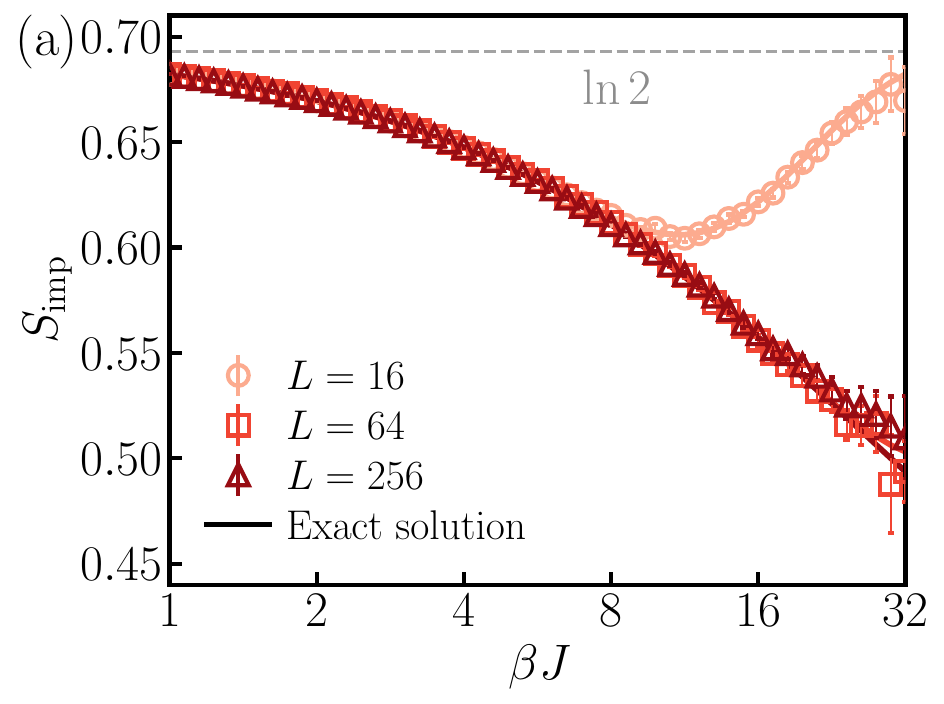}}
\hfill
\subfigure{\includegraphics[width=0.2385\textwidth]{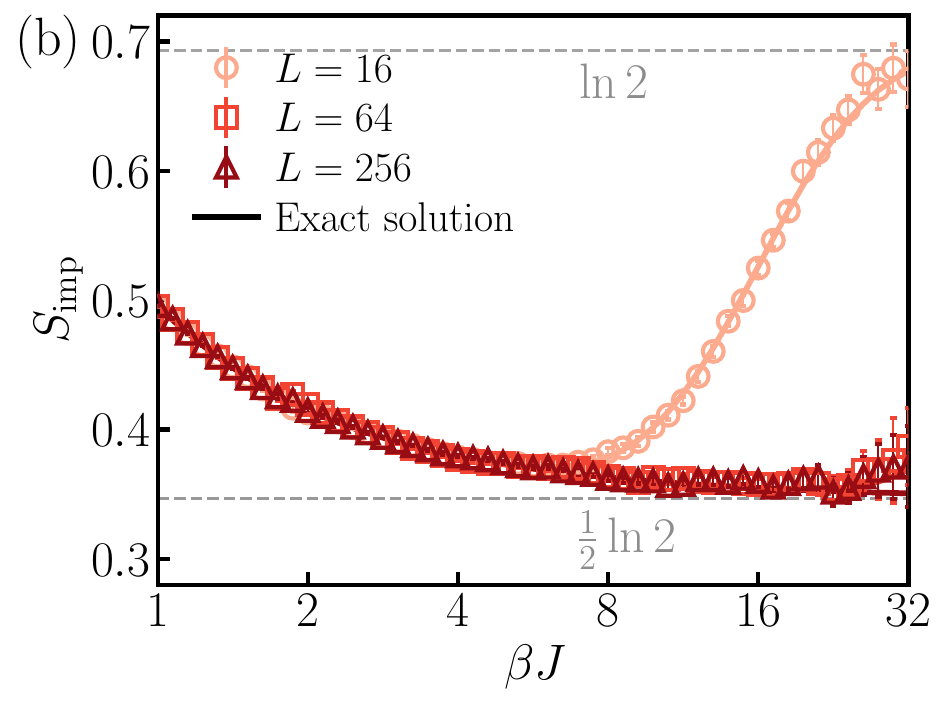}}
\caption{\justifying
Impurity entropy $S_{\mathrm{imp}}$ as a function of $\beta J$ for the critical TFIM with (a) $\lambda=0.2$ and (b) $\lambda=1.0$. 
The solid lines show exact free fermion solutions for different bath lengths $L$~\cite{SupMat}.
The horizontal dashed lines mark $\ln 2$ and $\frac{1}{2}\ln 2$.
}
\label{fig:TFIM}
\end{figure}

Fig.~\ref{fig:TFIM} compares the Local-Maxwell results with the exact free fermion solutions~\cite{SupMat}. 
At the weak coupling $\lambda=0.2$, $S_{\mathrm{imp}}$ of larger systems decreases gradually upon cooling, without reaching the boundary-entropy plateau within the temperature range studied.
In contrast, for $\lambda=1.0$, the larger systems approach the universal boundary-entropy value $\frac{1}{2}\ln 2$, 
associated with the free-to-fixed boundary renormalization group flow of the critical Ising model~\cite{PhysRevLett.67.161,Chatterjee1996}. 
For both couplings, the smallest system shows a low-temperature upturn toward $\ln2$. 
This follows from the twofold ground-state degeneracy of the finite system, since the impurity spin has no transverse field.
The Local-Maxwell results reproduce these distinct temperature dependences and finite-size effects, 
confirming the accuracy of the entropy reconstruction across different couplings and system sizes.

\paragraph{Interacting XXZ spin chain.---}
We next examine the statistical efficiency of the Local-Maxwell method in a spin-$1/2$ XXZ chain with a boundary impurity~\cite{Affleck:1992prb,PhysRevB.58.5529}. 
The Hamiltonian reads
\begin{align} 
	\mathcal{H}_{\text{XXZ}} &= -\sum_{j=1}^{L-1}\left[J_x(S_j^xS_{j+1}^x+S_j^yS_{j+1}^y)+J_zS_j^zS_{j+1}^z\right] \nonumber\\
	& \ \ \ - C_x(S_0^xS_1^x+S_0^yS_1^y) - C_zS_0^zS_1^z,
	\label{eq:Hxxz}
\end{align}
where $S_0$ denotes the impurity spin-$1/2$, $L$ is the number of bath spins, and $C_x$ and $C_z$ are the transverse and longitudinal impurity couplings. 
We choose $J_x=-1.25$, $J_z=-1$, corresponding to a gapless Luttinger liquid bath with $\Delta=J_z/J_x=0.8$, while the impurity couplings are set as $C_x=0.2$ (or $1.0$) and $C_z=0$. 
%
\begin{figure}[tbp]
\centering
\subfigure{\includegraphics[width=0.2385\textwidth]{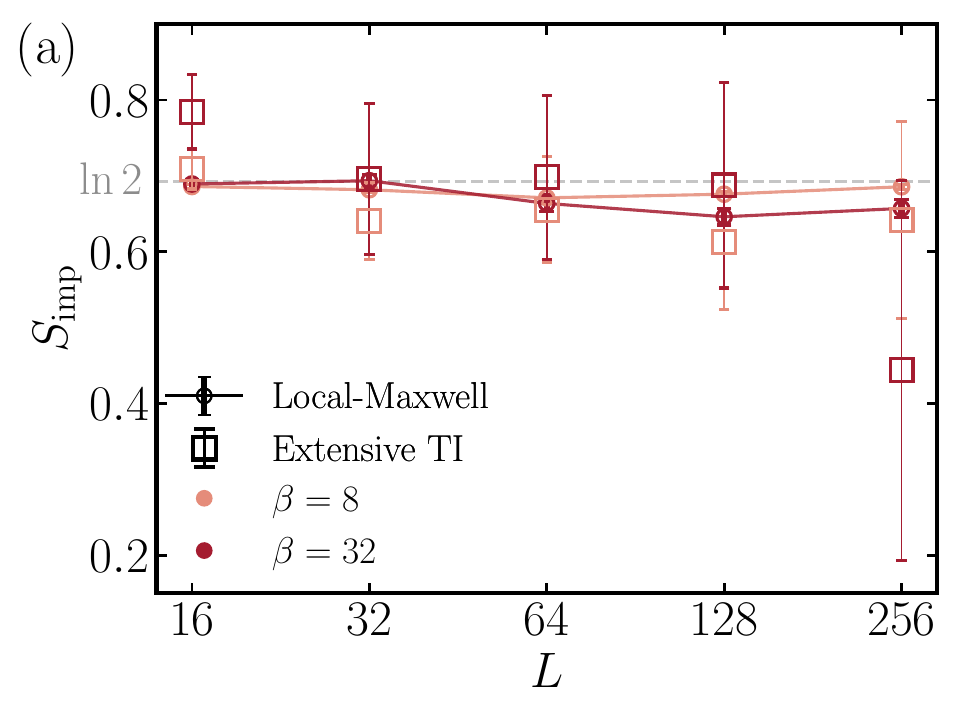}}
\hfill
\subfigure{\includegraphics[width=0.2385\textwidth]{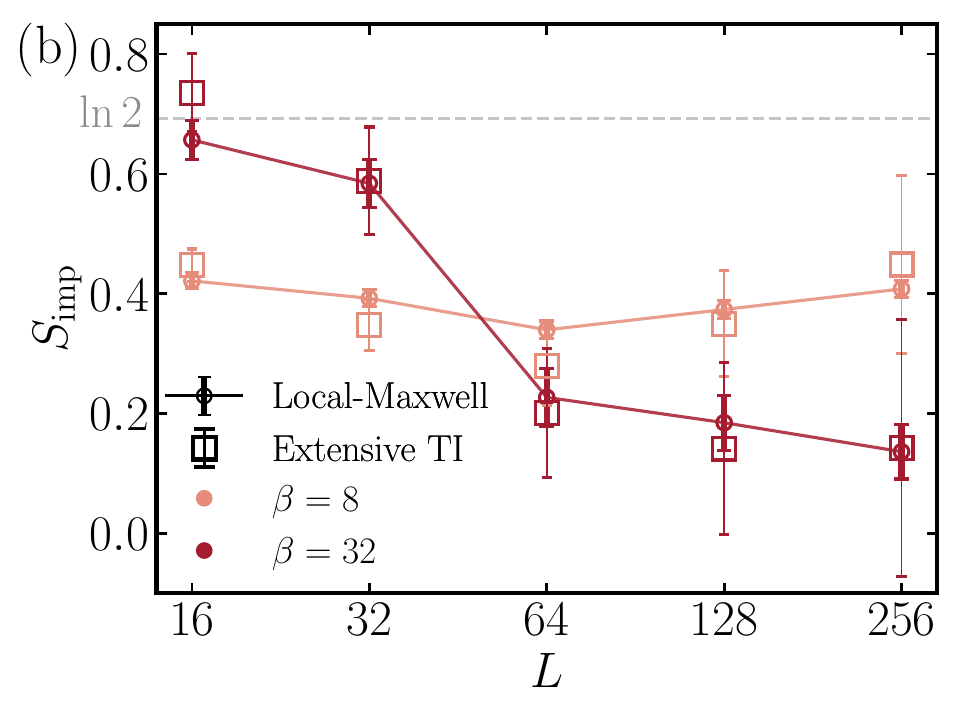}}
\hfill
\subfigure{\includegraphics[width=0.2385\textwidth]{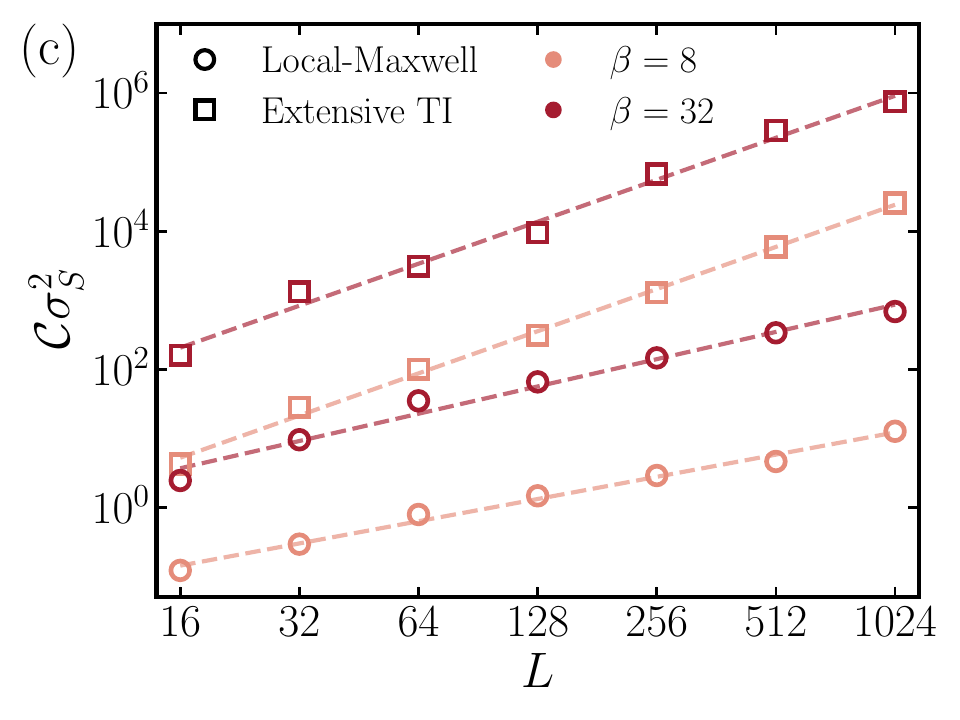}}
\hfill
\subfigure{\includegraphics[width=0.2385\textwidth]{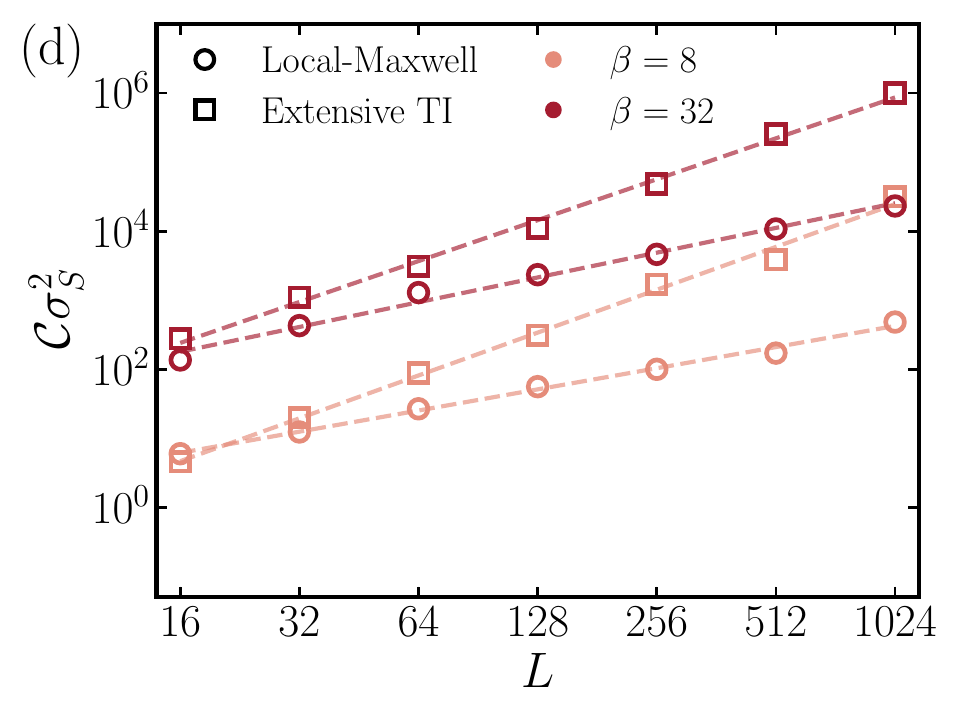}}
\caption{\justifying
Impurity entropy and statistical efficiency in the XXZ spin chain. 
(a,b) Impurity entropy $S_{\mathrm{imp}}$ as a function of bath length $L$ for (a) $C_x=0.2$ and (b) $C_x=1.0$, 
comparing Local-Maxwell and extensive TI methods at $\beta=8$ and $32$.
(c,d) Corresponding cost-weighted variances $\mathcal{V}$ for (c) $C_x=0.2$ and (d) $C_x=1.0$.  
Dashed lines indicate power-law fits over $16\leq L\leq1024$.
}
\label{fig:XXZ}
\end{figure}

Fig.~\ref{fig:XXZ} (a) and (b) compare the impurity entropy obtained using the Local-Maxwell method with extensive TI, 
in which the entropies of the full system and the bare bath are obtained separately from their extensive energies. 
Both methods give consistent results, while Local-Maxwell exhibits substantially smaller statistical errors, especially at larger system sizes. 
At weak coupling $C_x=0.2$, the impurity entropy remains close to $\ln2$, with only a modest reduction upon cooling.
On the contrary, for stronger coupling $C_x=1.0$, the impurity entropy is substantially suppressed and shows a larger finite-size dependence at $\beta=32$.

To quantify their statistical efficiency, we introduce the cost-weighted variance $\mathcal{V} = \mathcal{C} \sigma_S^2$~\cite{SupMat}, 
where $\mathcal{C}$ and $\sigma_S$ denote the total computational cost and the standard error of the impurity entropy, respectively. 
As shown in Fig.~\ref{fig:XXZ} (c) and (d), power-law fits $\mathcal{V}=AL^p$ over all sizes yield an approximately quadratic size dependence for extensive TI, 
compared with $p\simeq1.0\sim1.3$ for the Local-Maxwell method.
Consequently, the relative statistical efficiency of Local-Maxwell increases with system size, consistent with the direct comparison shown in the Supplemental Material~\cite{SupMat}.
The improvement is particularly striking at weak impurity coupling, where the impurity contribution is small compared with the extensive bath energies.  
These results demonstrate the computational advantage of reconstructing impurity entropy from local statistical observables in interacting quantum systems.

\paragraph{Application to multichannel spin chain junction.---}
Multichannel quantum impurity systems exhibit rich boundary physics~\cite{nozieres1980kondo,AFFLECK1991641}, 
with impurity entropy providing a probe of their low-energy behavior.
We therefore consider a spin-$1/2$ impurity symmetrically coupled to the ends of three XXZ chains~\cite{PhysRevLett.128.146803,Gaines:2026aa}.
The Hamiltonian is
\begin{align}
\mathcal{H}=&J\sum_{\alpha=1}^{3}\sum_{j=1}^{L-1}\left(S_{\alpha,j}^{x}S_{\alpha,j+1}^{x}+S_{\alpha,j}^{y}S_{\alpha,j+1}^{y}+\Delta S_{\alpha,j}^{z}S_{\alpha,j+1}^{z}\right)\nonumber\\
&+J_K\sum_{\alpha=1}^{3}\left(S_d^xS_{\alpha,1}^x+S_d^yS_{\alpha,1}^y+\Delta S_d^zS_{\alpha,1}^z\right),
\label{eq:HM3xxz}
\end{align}
Here $J$ and $J_K$ are the intrachain and impurity exchange couplings, respectively, and $\Delta$ is the XXZ anisotropy. 
We set $J=1$ and $\Delta=0.8$. 
Each chain contains $L$ spins, giving $N=3L+1$ spins in the junction. 
We define $S_{\mathrm{imp}}=S_{\mathrm{junction}}-3S_{\mathrm{chain}}$, with the exact decoupled limit reference $S_{\mathrm{imp}}(J_K=0)=\ln 2$.

At $\Delta=0.8$, each chain is a gapless Luttinger liquid with parameter $K=\pi/[2(\pi-\arccos\Delta)]\simeq0.629$. 
The transverse impurity exchange has scaling dimension $d_\perp=1/(2K)\simeq0.795<1$ and is relevant at weak coupling. 
Within the weak coupling regime, boundary perturbation theory gives 
$S_{\mathrm{imp}}-\ln2\sim-AJ_K^2T^{2d_\perp-2}$ with $A > 0$, 
which predicts entropy depletion upon cooling~\cite{SupMat,affleck1995,cardy2008bCFT}.
At the asymptotic limit $J_K \gg J$, the impurity and the three chain endpoints form a four-spin cluster with a nondegenerate ground state. 
The remaining low-energy degrees of freedom form three shortened XXZ chains.
In this limit, $S_{\mathrm{imp}}\simeq-\pi T/v$ in the low-temperature thermodynamic regime, where $v$ is the spin velocity~\cite{SupMat}. 
The two limits favor opposite thermal responses.
%
\begin{figure}[tbp]
\centering
\subfigure{\includegraphics[width=0.2385\textwidth]{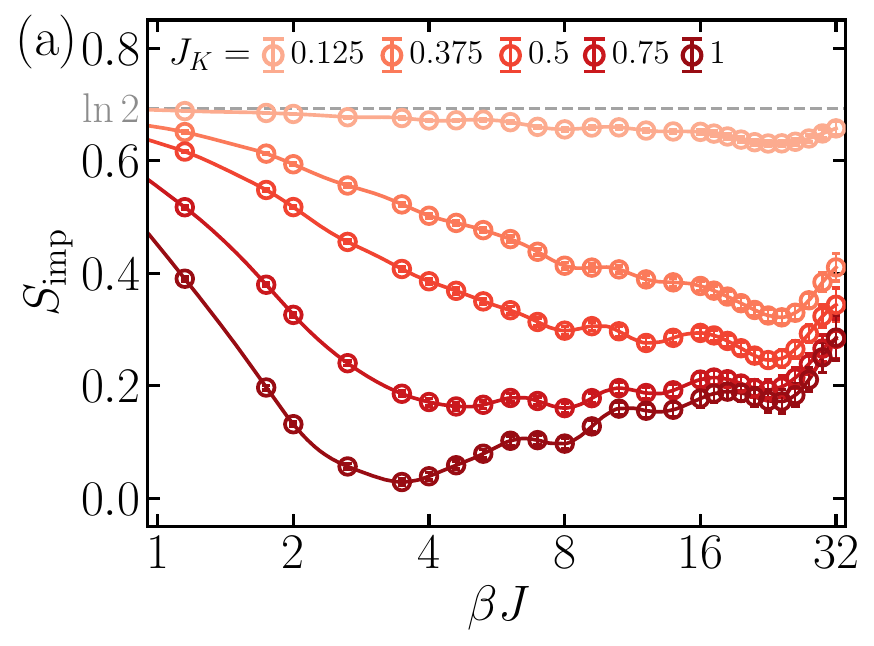}}
\hfill
\subfigure{\includegraphics[width=0.2385\textwidth]{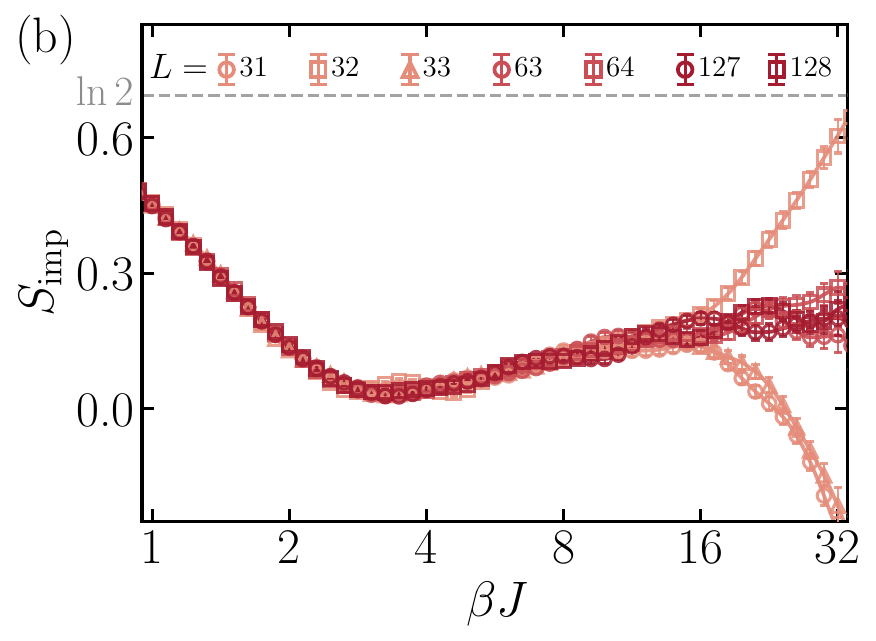}}
\hfill
\subfigure{\includegraphics[width=0.2385\textwidth]{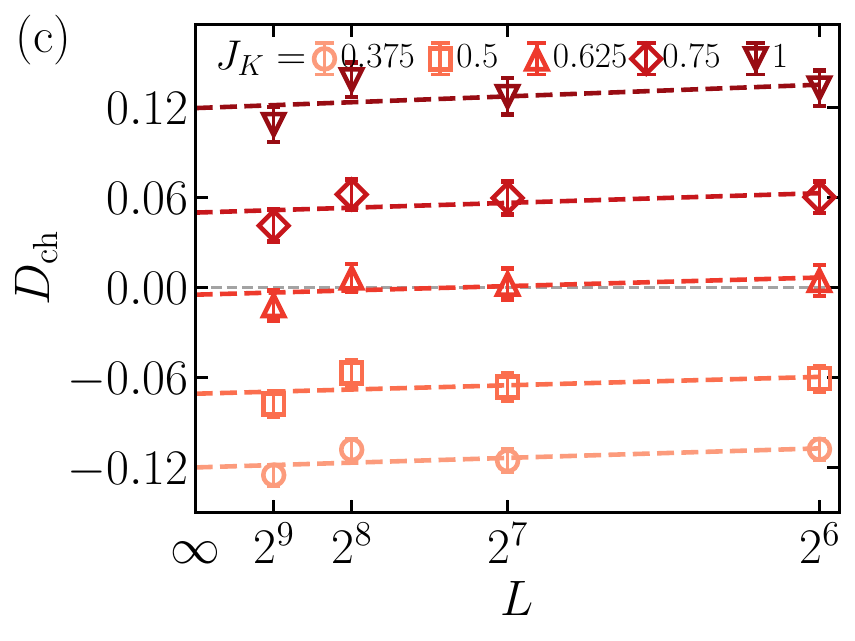}}
\hfill
\subfigure{\includegraphics[width=0.2385\textwidth]{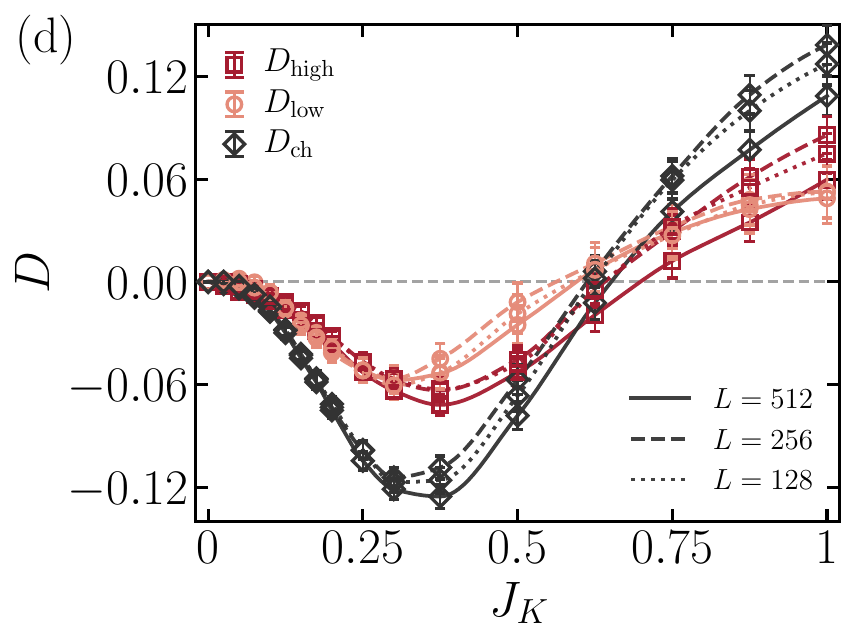}}
\caption{\justifying
Impurity entropy and thermal response of the three-channel XXZ model at $\Delta=0.8$.
(a) Impurity entropy $S_{\mathrm{imp}}$ as a function of $\beta J$ for $L=512$ and different $J_K$. 
Dashed line indicates the decoupled-impurity entropy $\ln2$.
(b) Temperature dependence of $S_{\mathrm{imp}}$ for neighboring odd and even chain lengths.
(c) System-size dependence of the free energy chord contrast 
$D_{\mathrm{ch}}$ for different $J_K$.
Dashed lines represent linear fits in $1/L$.
(d) Low- and high-temperature contributions, $D_{\mathrm{low}}$ and $D_{\mathrm{high}}$, to the total chord contrast $D_{\mathrm{ch}}=D_{\mathrm{low}}+D_{\mathrm{high}}$.
}
\label{fig:M3XXZ}
\end{figure}

We compute the impurity entropy for the intermediate region $0<J_K\leq1$ using the Local-Maxwell method. 
At high temperatures, the results agree with the expansion $S_{\mathrm{imp}}=\ln2-3(2+\Delta^2)(\beta J_K)^2/32+\mathcal O(\beta^3)$, derived in the Supplemental Material~\cite{SupMat}.
Fig.~\ref{fig:M3XXZ}(a) shows entropy depletion upon cooling at weak coupling.
At larger $J_K$, $S_{\mathrm{imp}}$ reaches a minimum and increases upon further cooling. 
We further examine whether the entropy recovery arises from finite-size parity effects. 
For even $L$, the junction contains an odd number of spin-$1/2$ degrees of freedom and exhibits Kramers degeneracy~\cite{tasaki2020physics}. 
By contrast, Kramers degeneracy is instead enforced in each reference chain for odd $L$.
Fig.~\ref{fig:M3XXZ}(b) shows strong parity effects in the low-temperature impurity entropy of smaller systems. 
The recovery observed for both $L=127$ and $128$ rules out a purely parity-induced origin.

To characterize the entropy recovery without numerical temperature differentiation, we introduce the free energy chord entropy
\begin{equation}
\begin{aligned}
S_{\mathrm{ch}}(T_1,T_2)
&=-\frac{F_{\mathrm{imp}}(T_2)-F_{\mathrm{imp}}(T_1)}{T_2-T_1}\\
&=\frac{1}{T_2-T_1}\int_{T_1}^{T_2}S_{\mathrm{imp}}(T)\,\mathrm{d}T,
\end{aligned}
\label{eq:M3chord}
\end{equation}
which represents the temperature averaged impurity entropy over $[T_1,T_2]$.
The impurity free energy is obtained by integrating the local conjugate observable over $J_K$~\cite{SupMat}.
We define the cold-minus-warm chord contrast as
\begin{equation}
D_{\mathrm{ch}}(J_K,L)=S_{\mathrm{ch}}(1/32,1/16)-S_{\mathrm{ch}}(1/8,1/4).
\label{eq:M3contrast}
\end{equation}
Negative and positive values of $D_{\mathrm{ch}}$ indicate entropy depletion and recovery upon cooling, respectively.
As shown in Fig.~\ref{fig:M3XXZ}(c), $D_{\mathrm{ch}}$ changes sign with increasing $J_K$. 
Linear fits in $1/L$ suggest that this reversal persists in the thermodynamic limit.

The coupling dependence of $D_{\mathrm{ch}}$ can be further resolved by comparing adjacent temperature windows.
Let $S_{\mathrm c}$, $S_{\mathrm m}$, and $S_{\mathrm w}$ denote the chord entropies over $[1/32,1/16]$, $[1/16,1/8]$, and $[1/8,1/4]$, respectively.
We define $D_{\mathrm{low}}=S_{\mathrm c}-S_{\mathrm m}$ and $D_{\mathrm{high}}=S_{\mathrm m}-S_{\mathrm w}$, 
such that $D_{\mathrm{ch}}=D_{\mathrm{low}}+D_{\mathrm{high}}$.
As shown in Fig.~\ref{fig:M3XXZ}(d), both contrasts are negative at weak coupling but become positive at larger $J_K$.
The lower temperature contrast tends to change sign first, suggesting that 
entropy recovery develops at lower temperatures before extending to the higher temperature window.
This evolution reveals a coupling dependent reversal of the finite-temperature impurity entropy response.

The nonmonotonic impurity entropy has also been reported in boundary spin chain models, 
where localized bound modes reorganize the low-energy excitation spectrum~\cite{Zhakenov2025,Zhakenov2026}. 
Unlike conventional local impurity singlet screening, the strong coupling limit of our three-arm junction yields a nondegenerate four-spin cluster ground state derived from a spin triplet~\cite{SupMat}.
%
The opposite thermal responses in the weak and strong coupling limits suggest that a reorganization of thermally accessible boundary excitations may underlie the observed entropy recovery. 
The Local-Maxwell reconstruction and free energy chord analysis reveal a coupling dependent reversal of the finite-temperature impurity entropy response, 
consistent with a boundary crossover.

\paragraph{Conclusions and discussions.---}
In this work, we have introduced a Local-Maxwell method that reconstructs impurity entropy from local coupling responses in worldline QMC.
For nonnegative worldline expansions with bounded local coupling operators, we establish bath-size independent fluctuation bounds for local vertex counts and, 
under a spin-flip parity condition, for the corresponding Maxwell estimator.
The TFIM benchmark validates the reconstruction against exact results, 
while the interacting XXZ chain demonstrates improved statistical efficiency over conventional extensive thermodynamic integration. 
In the three-channel XXZ junction, the method reveals nonmonotonic impurity entropy upon cooling and a coupling dependent reversal of the cold-minus-warm entropy contrast.
%
Our framework offers a practical route to impurity and defect thermodynamics in interacting spin chains~\cite{PhysRevB.48.9528,PhysRevB.96.060403}, ladders~\cite{PhysRevB.65.144411}, 
and higher dimensional quantum magnets~\cite{PhysRevB.70.024406,PhysRevLett.99.027205,HUANG202341}, 
as well as Kondo type systems~\cite{Kattel2026PRB,Paris2026PRL:aa,Paris2026PRL:ab,Gaines:2026aa} 
where local coupling responses can be sampled efficiently.
 
\begin{acknowledgments}
\paragraph{Acknowledgments.---}
The author thanks 
Yuan Yao and Ji Zou for their support, 
Tengzhou Zhang and Tianwei Ouyang for helpful discussions on QMC implementations and the use of computing clusters, 
and Haojie Shen, Tu Hong and Zenan Dai for useful discussions. 
\end{acknowledgments}


\bibliographystyle{apsrev4-1}

%

\section*{End Matter}
\paragraph{Bounds on local vertex count moments.---}
We establish bath size \textit{independent} bounds on all factorial moments of the local impurity vertex count $N_D$, from which a variance bound follows.
Consider $\mathcal{H}(\lambda)=\mathcal{H}_0+\lambda V_D$, with $\|V_D\|\leq v_D$, 
where $v_D$ is independent of the bath size.
We assume the nonnegative worldline expansion in Eq.~\eqref{eq:local_partition} for $\lambda\geq0$, with all other Hamiltonian parameters held fixed. 
For integer $k\geq1$, the factorial moments satisfy
\begin{equation}
\langle(N_D)_k\rangle=\frac{\lambda^k}{\mathcal Z}\frac{\partial^k \mathcal Z}{\partial\lambda^k},
\end{equation}
where $(N_D)_k=N_D(N_D-1)\cdots(N_D-k+1)$ and $(N_D)_0=1$. 
We note that the nonnegative configuration weights ensure $\langle(N_D)_k\rangle\geq0$. 

Since $\mathcal{H}$ and $V_D$ generally do not commute, 
the $k$-th derivative of the partition function is expressed through the Duhamel expansion~\cite{bhatia2009positive} as
\begin{align}
\frac{\partial^k \mathcal Z}{\partial\lambda^k}
=&(-\beta)^k k!\int_{\Delta_k}\mathrm{d}^k s\nonumber\\
&\times\operatorname{Tr}\left[\mathrm{e}^{-\beta t_0 \mathcal H}V_D \mathrm{e}^{-\beta t_1 \mathcal H}\cdots V_D \mathrm{e}^{-\beta t_k \mathcal H}\right],
\end{align}
where $\Delta_k=\{0\leq s_k\leq\cdots\leq s_1\leq1\}$, $t_0=1-s_1$, $t_j=s_j-s_{j+1}$ for $1\leq j<k$, and $t_k=s_k$.
These intervals satisfy $t_j\geq0$ and $\sum_{j=0}^k t_j=1$.
By applying the generalized Schatten-H\"older inequality~\cite{bhatia2013matrix,schatten2013norm}, 
with exponent $1/t_j$ for each imaginary-time propagator and the operator norm for each $V_D$, 
one obtains
\begin{align}
&\left|\operatorname{Tr}\left[\mathrm{e}^{-\beta t_0 \mathcal H}V_D\cdots V_D \mathrm{e}^{-\beta t_k \mathcal H}\right]\right|\nonumber\\
&\qquad\leq\|V_D\|^k\prod_{j=0}^k\left\|\mathrm{e}^{-\beta t_j \mathcal H}\right\|_{1/t_j}
=\mathcal Z\|V_D\|^k.
\end{align}
%
For $t_j>0$, the Schatten norm satisfies
$\|e^{-\beta t_j\mathcal H}\|_{1/t_j}
=[\operatorname{Tr}(e^{-\beta t_j\mathcal H})^{1/t_j}]^{t_j}
=\mathcal Z^{t_j}$.
At $t_j=0$, the corresponding operator norm is
$\|\mathbb I\|_\infty=1=\mathcal Z^0$.
Since the sum of $t_j$ equals $1$, the product of propagator norms equals $\mathcal Z$, i.e.,
\begin{equation}
\left|\frac{\partial^k \mathcal Z}{\partial\lambda^k}\right|\leq\beta^k \mathcal Z\|V_D\|^k.
\end{equation}
Combining this with the factorial moment identity and positivity of the worldline weights, we obtain
\begin{equation}
0\leq\langle(N_D)_k\rangle\leq b^k,\qquad b=\beta\lambda v_D.
\end{equation}
This bound holds for every integer $k\geq1$ and is independent of the bath size at fixed $\beta$ and $\lambda$.
It requires neither a spectral gap nor a finite correlation length.

\paragraph{Statistical implications.---}
The first two factorial-moment bounds give $\langle N_D\rangle\leq b$ and $\langle(N_D)_2\rangle\leq b^2$.
Using $N_D^2=(N_D)_2+N_D$, we obtain
\begin{equation}
\operatorname{Var}(N_D)
=\langle(N_D)_2\rangle+\langle N_D\rangle-\langle N_D\rangle^2
\leq b+b^2,
\end{equation}
where $\operatorname{Var}(N_D)$ denotes the variance $\operatorname{Var}(N_D) = \langle N_D^2\rangle - \langle N_D \rangle^2$.
The all order bound also controls higher ordinary moments, which are finite linear combinations of factorial moments. 
%
%

For nonnegative worldline representations with bounded local diagonal interactions, 
a corresponding bath-size independent variance bound extends to the full estimator $Q_\lambda$ at fixed $\beta$ and $\lambda>0$, as shown in the Supplemental Material~\cite{SupMat}. 

\clearpage
\onecolumngrid 
\appendix
\setcounter{equation}{0}
\renewcommand\theequation{S\arabic{equation}}
\setcounter{figure}{0}
\renewcommand{\thefigure}{S\arabic{figure}}
\setcounter{table}{0}
\renewcommand{\thetable}{S\arabic{table}}
\section{\Large{} Supplemental Materials of ``Local Thermodynamic Reconstruction of Impurity Entropy''}

\subsection{Local estimator fluctuation bounds}
In this section, we extend the factorial moment bounds derived in the End Matter to local estimators that include diagonal interactions. 
We then apply these bounds to the transverse-field Ising model (TFIM), the XXZ spin chain, and the three-channel XXZ impurity model studied in the main text.

\subsubsection{Unified treatment of local couplings}
We consider local impurity couplings that may include both off-diagonal exchange and diagonal interactions in the worldline basis.
To establish a unified variance bound, we introduce two independent coupling parameters through
\begin{equation}
\mathcal H(g,h)=\mathcal H_0+gV_\perp+hV_z,
\label{eq:general_hamiltonian_sm}
\end{equation}
where $V_\perp$ generates exchange vertices and $V_z$ is diagonal in the worldline basis.
Both operators are local and bounded.

Assume a nonnegative worldline expansion with configuration weights
\begin{equation}
W(\mathscr C;g,h,\beta)=g^{N_D(\mathscr C)}\mathrm{e}^{-\beta hA_D(\mathscr C)}\bar W(\mathscr C,\beta),
\label{eq:general_weight_sm}
\end{equation}
where $g\geq0$ and $\bar W(\mathscr C,\beta)\geq0$ is independent of $g$ and $h$.
The diagonal observable is defined by
\begin{equation}
A_D(\mathscr C)=\frac{1}{\beta}\int_0^\beta\mathrm{d}\tau\,V_z(\tau;\mathscr C),
\label{eq:diagonal_observable_sm}
\end{equation}
where $V_z(\tau;\mathscr C)$ denotes the diagonal matrix element along the sampled worldline. 
Since $V_z$ is bounded, $|A_D(\mathscr C)|\leq v_z$, with $v_z=\|V_z\|$ independent of the bath size.

Along a physical coupling path $g=\lambda$ and $h=h(\lambda)$, we define the configuration weight
$\widetilde W(\mathscr C;\lambda,\beta)
=W(\mathscr C;\lambda,h(\lambda),\beta)$.
The local estimator is then
\begin{equation}
Q_\lambda(\mathscr C)=
\frac{\partial}{\partial\lambda}
\ln\widetilde W(\mathscr C;\lambda,\beta)
=\frac{N_D(\mathscr C)}{\lambda}
-\beta h'(\lambda)A_D(\mathscr C).
\label{eq:general_Q_sm}
\end{equation}
where $h'(\lambda)=\mathrm{d}h/\mathrm{d}\lambda$ and $\lambda>0$.
Its expectation satisfies $\langle Q_\lambda\rangle=\mathrm{d}\ln\mathcal Z/\mathrm{d}\lambda$ along the same coupling path.

For observables $X$ and $Y$, one can define the variance and covariance as
\begin{equation}
\operatorname{Var}(X)=\langle X^2\rangle-\langle X\rangle^2,\qquad
\operatorname{Cov}(X,Y)=\langle XY\rangle-\langle X\rangle\langle Y\rangle.
\label{eq:variance_covariance_sm}
\end{equation}
To bound the fluctuations of $N_D$, we treat $g$ and $h$ as independent parameters and differentiate the partition function with respect to $g$ at fixed $h$.
Denoting the thermal average at fixed $g$ and $h$ by $\langle\cdots\rangle_{g,h}$, the all order factorial moment bound established in the End Matter gives
\begin{equation}
0\leq\langle(N_D)_k\rangle_{g,h}\leq b_\perp^k,\qquad b_\perp=\beta g\|V_\perp\|.
\label{eq:ND_moment_general_sm}
\end{equation}
Evaluating this bound along the physical coupling path, where $g=\lambda$ and $b_\perp=\beta\lambda\|V_\perp\|$, the first two factorial moments yield
\begin{equation}
\operatorname{Var}(N_D)\leq b_\perp+b_\perp^2.
\label{eq:ND_variance_general_sm}
\end{equation}

The variance of the full estimator is
\begin{equation}
\operatorname{Var}(Q_\lambda)=\frac{\operatorname{Var}(N_D)}{\lambda^2}+\beta^2[h'(\lambda)]^2\operatorname{Var}(A_D)-\frac{2\beta h'(\lambda)}{\lambda}\operatorname{Cov}(N_D,A_D).
\label{eq:Q_variance_decomposition_sm}
\end{equation}
Using $\operatorname{Var}(A_D)\leq v_z^2$ and applying the Cauchy-Schwarz inequality to the covariance term, we obtain
\begin{equation}
\operatorname{Var}(Q_\lambda)\leq\left[\frac{\sqrt{\operatorname{Var}(N_D)}}{\lambda}+\beta|h'(\lambda)|\sqrt{\operatorname{Var}(A_D)}\right]^2\leq\left[\frac{\sqrt{b_\perp+b_\perp^2}}{\lambda}+\beta|h'(\lambda)|v_z\right]^2.
\label{eq:Q_variance_bound_sm}
\end{equation}
For fixed $\beta$ and $\lambda>0$, with $h'(\lambda)$ independent of the bath size, 
this bound is bath-size independent and accounts for correlations between transverse and diagonal contributions.

\subsubsection{Application of fluctuation bounds to the quantum spin models}
%
The resulting bounds can be evaluated explicitly for the three models studied in the main text. 
For the TFIM, the impurity coupling operator is
\begin{equation}
V_{\mathrm{TFIM}}=-\sigma_0^z\sigma_1^z,\qquad
\|V_{\mathrm{TFIM}}\|=1.
\end{equation}
In the vertex-count representation, $Q_\lambda=N_D/\lambda$.
The resulting estimator variance is bounded by
\begin{equation}
\operatorname{Var}(Q_\lambda)\leq\frac{\beta}{\lambda}+\beta^2.
\end{equation}

For the XXZ benchmark with $C_z=0$, the local coupling operator is
\begin{equation}
V_{\mathrm{XXZ}}=-(S_0^xS_1^x+S_0^yS_1^y),\qquad
\|V_{\mathrm{XXZ}}\|=\frac12.
\end{equation}
Since $Q_{C_x}=N_D/C_x$, the estimator variance satisfies
\begin{equation}
\operatorname{Var}(Q_{C_x})\leq\frac{\beta}{2C_x}+\frac{\beta^2}{4}.
\end{equation}

For the three-channel XXZ model, the physical coupling path is $g=J_K$ and $h(J_K)=\Delta J_K$, with $h'(J_K)=\Delta$.
The transverse and longitudinal coupling operators are
\begin{equation}
V_\perp=\sum_{\alpha=1}^3(S_d^xS_{\alpha,1}^x+S_d^yS_{\alpha,1}^y),\qquad
V_z=\sum_{\alpha=1}^3S_d^zS_{\alpha,1}^z.
\end{equation}
Here, $N_D$ counts the transverse exchange vertices, while $A_D(\mathscr C)=\beta^{-1}\int_0^\beta\mathrm{d}\tau\,V_z(\tau;\mathscr C)$ is the imaginary-time average of the longitudinal exchange interaction.
The local estimator is therefore
\begin{equation}
Q_{J_K}(\mathscr C)=\frac{N_D(\mathscr C)}{J_K}-\beta\Delta A_D(\mathscr C).
\end{equation}
To evaluate the operator norms, we introduce the total spin $\mathbf T=\sum_{\alpha=1}^3\mathbf S_{\alpha,1}$ of the three chain endpoints, giving
\begin{equation}
V_\perp=\frac12(S_d^+T^-+S_d^-T^+),\qquad V_z=S_d^zT^z.
\label{eq:M3XXZ_T_sm}
\end{equation}
The endpoint-spin Hilbert space decomposes into one $T=3/2$ sector and two $T=1/2$ sectors.
In a fixed-$T$ sector, the largest matrix element of $V_\perp$ is $\frac12\sqrt{T(T+1)+\frac14}$, while the largest eigenvalue magnitude of $V_z$ is $T/2$.
Consequently,
\begin{align}
&\|V_\perp\|_{T=1/2}=\frac12,\qquad
\|V_\perp\|_{T=3/2}=1,\\
&\|V_z\|_{T=1/2}=\frac14,\qquad
\|V_z\|_{T=3/2}=\frac34.
\end{align}
The full operator norms are thus $\|V_\perp\|=1$ and $\|V_z\|=3/4$.
Substituting these values into Eq.~\eqref{eq:Q_variance_bound_sm}, we obtain
\begin{equation}
\operatorname{Var}(Q_{J_K})\leq
\left[
\frac{\sqrt{\beta J_K+(\beta J_K)^2}}{J_K}
+\frac34\beta|\Delta|
\right]^2.
\end{equation}

When the impurity exchange vertices occur only in even numbers, $N_D^2\leq2(N_D)_2$ gives
\begin{equation}
\left\langle\left(\frac{N_D}{g}\right)^2\right\rangle\leq2\beta^2\|V_\perp\|^2.
\end{equation}
Using $|A_D|\leq\|V_z\|$, we obtain
\begin{equation}
\operatorname{Var}(Q_g)\leq\beta^2\left(\sqrt{2}\|V_\perp\|+|h'(g)|\|V_z\|\right)^2.
\label{eq:Q_variance_even_sm}
\end{equation}

For the TFIM, the even vertex count condition gives
\begin{equation}
\operatorname{Var}(Q_\lambda)\leq2\beta^2.
\end{equation}
%
For the XXZ benchmark, the same condition yields
\begin{equation}
\operatorname{Var}(Q_{C_x})\leq\frac{\beta^2}{2}.
\end{equation}
%
Substituting these operator norms into Eq.~\eqref{eq:Q_variance_even_sm}, we obtain
\begin{equation}
\operatorname{Var}(Q_{J_K})\leq\beta^2\left(\sqrt{2}+\frac{3|\Delta|}{4}\right)^2.
\end{equation}
The bound remains independent of the bath length and uniform as $J_K\to0^+$.

\subsection{Transverse-field Ising model}
%
This section presents the exact finite-size free fermion solution of the TFIM, which provides the solid curves in Fig.~1 of the main text. 
We also present additional Local-Maxwell QMC results extending the temperature range and system-size coverage of the benchmark.

\subsubsection{Exact free fermion solution}
We start from the TFIM Hamiltonian 
\begin{equation}
  \mathcal{H}_{\text{TFIM}} = \mathcal{H}_{\text{TFIM}}^{(\text{bath})} + \mathcal{H}_{\text{TFIM}}^{(\text{imp})},
\end{equation}
where 
\begin{align}
  \mathcal{H}_{\text{TFIM}}^{(\text{bath})} &= -J \sum_{j=1}^{L-1} \sigma^z_j \sigma^z_{j+1} - \Gamma \sum_{j=1}^{L} \sigma^x_j, \\
  \mathcal{H}_{\text{TFIM}}^{(\text{imp})} &= -\lambda \sigma^z_0 \sigma^z_{1},
\end{align}
where $L$ denotes the number of bath spins, and $\sigma_0$ represents the impurity spin without a transverse field. 
We take $J=\Gamma=1$ here, corresponding to the quantum critical point of the bath. 

The model can be mapped onto free fermions through the Jordan-Wigner transformation. 
For convenience, we first perform a uniform rotation of the spin axes, $\sigma_j^z\rightarrow\sigma_j^x$ and $\sigma_j^x\rightarrow\sigma_j^z$. 
The Hamiltonian then becomes 
\begin{equation}
  \mathcal{H}_{\text{TFIM}} = -\sum_{j=0}^{L-1}K_j\sigma_j^x\sigma_{j+1}^x -\sum_{j=0}^{L}h_j\sigma_j^z,
\end{equation}
where $K_0=\lambda$ and $h_0=0$, $K_j=J$ and $h_j=\Gamma$ for $j\geq1$. 

By introducing the Majorana operators
\begin{equation}
  a_j= \left(\prod_{k=0}^{j-1}\sigma_k^z\right)\sigma_j^x, \ \ \ 
  b_j= \left(\prod_{k=0}^{j-1}\sigma_k^z\right)\sigma_j^y,
\end{equation}
which satisfy $\{a_i,a_j\}=\{b_i,b_j\}=2\delta_{ij}$ and $\{a_i,b_j\}=0$, 
we can obtain the quadratic Hamiltonian
\begin{equation}
  \mathcal{H}_{\mathrm{TFIM}}=\mathrm{i}\sum_{j=0}^{L-1}K_jb_ja_{j+1} + \mathrm{i}\sum_{j=0}^{L}h_j a_jb_j
  =\mathrm{i}\bm{a}^{\mathsf{T}}\bm{B}\bm{b},
\end{equation}
where $\boldsymbol{a}=(a_0,\ldots,a_L)^{\mathsf{T}}$ and $\boldsymbol{b}=(b_0,\ldots,b_L)^{\mathsf{T}}$,
%
and the nonzero matrix elements of $\bm{B}$ are $B_{jj}=h_j$ and $B_{j+1,j}=-K_j$. 
The fermionic excitation energies are given by $\epsilon_n=2s_n$, where $s_n$ are the singular values of $\bm{B}$. 
Since $h_0=0$, the matrix $\bm{B}$ has an exact zero singular value, $\epsilon_0=0$.

To obtain the finite-size spectrum explicitly, we consider the nonzero eigenvalues of $\bm{BB}^{\mathsf{T}}$. 
Removing its zero row and column gives the $L\times L$ tridiagonal matrix
\begin{equation}
  \bm{M}_L=
  \begin{pmatrix}
    \Gamma^2+\lambda^2 & -\Gamma J & & \\
    -\Gamma J & \Gamma^2+J^2 & \ddots & \\
    & \ddots & \ddots & -\Gamma J\\
    & & -\Gamma J & \Gamma^2+J^2
  \end{pmatrix}.
\end{equation}
At the critical point $J=\Gamma=1$, the eigenvalue equation $\bm{M}_L\boldsymbol{v}=\mu\boldsymbol{v}$ yields the bulk recurrence relation
\begin{equation}
  -v_{j-1}+2v_j-v_{j+1}=\mu v_j,
\end{equation}
with the boundary conditions $v_{L+1}=0$ and $v_0=(1-\lambda^2)v_1$. 
By writing $\mu=2-2\cos k$ and $v_j\propto\sin[(L+1-j)k]$, we obtain the finite-size quantization condition
\begin{equation}
  \sin[(L+1)k]=(1-\lambda^2)\sin(Lk).
  \label{eq:TFIM_quantization}
\end{equation}

For the couplings considered here, the $L$ solutions $k_n\in(0,\pi)$ determine the nonzero excitation energies
\begin{equation}
  \epsilon_n(L,\lambda)=4\sin\frac{k_n}{2},\qquad n=1,\ldots,L.
\end{equation}
For the isolated bath, the corresponding quantization condition is obtained by setting $\lambda=0$, giving
\begin{equation}
  \widetilde{\epsilon}_n(L)=4\sin\frac{q_n}{2},\qquad 
  q_n=\frac{(2n-1)\pi}{2L+1}.
\end{equation}
Thus, both the full system and the isolated bath have exact finite-size single-particle spectra.

The diagonalized Hamiltonian is
\begin{equation}
  \mathcal{H}_{\mathrm{TFIM}}=\sum_{n=0}^{L}\epsilon_n\left(c_n^\dagger c_n-\frac{1}{2}\right),
\end{equation}
where $\epsilon_0=0$ represents the zero-energy fermionic mode. The corresponding partition function is
\begin{equation}
  \mathcal{Z}_{\mathrm{full}}=2\prod_{n=1}^{L}2\cosh\left(\frac{\beta\epsilon_n}{2}\right).
\end{equation}
Using $S=(1-\beta\partial_\beta)\ln \mathcal{Z}$, we define the entropy contribution of a single fermionic mode as
\begin{equation}
  s(\beta,\epsilon)=\ln\left[2\cosh\left(\frac{\beta\epsilon}{2}\right)\right]-\frac{\beta\epsilon}{2}\tanh\left(\frac{\beta\epsilon}{2}\right).
\end{equation}
The exact impurity entropy is then obtained by subtracting the entropy of the isolated bath from that of the full system,
\begin{equation}
  S_{\mathrm{imp}}^{\mathrm{exact}}(\beta,L,\lambda)=\ln2+\sum_{n=1}^{L}\left[s(\beta,\epsilon_n(L,\lambda))-s(\beta,\widetilde{\epsilon}_n(L))\right].
  \label{eq:TFIM_exact_entropy}
\end{equation}

Equation~\eqref{eq:TFIM_exact_entropy} provides the exact finite-size impurity entropy used to benchmark the Local-Maxwell QMC results in the main text. At $\lambda=0$, the impurity is completely decoupled, and $S_{\mathrm{imp}}=\ln2$ for arbitrary temperature. For any fixed finite $L$, the zero-energy fermionic mode gives a twofold ground-state degeneracy of the full system, yielding $S_{\mathrm{imp}}\rightarrow\ln2$ as $\beta\rightarrow\infty$. This accounts for the low-temperature upturn observed in small systems.

\subsubsection{Extended temperature and system-size results}
\begin{figure}[htbp]
\centering
\subfigure{\includegraphics[width=0.495\textwidth]{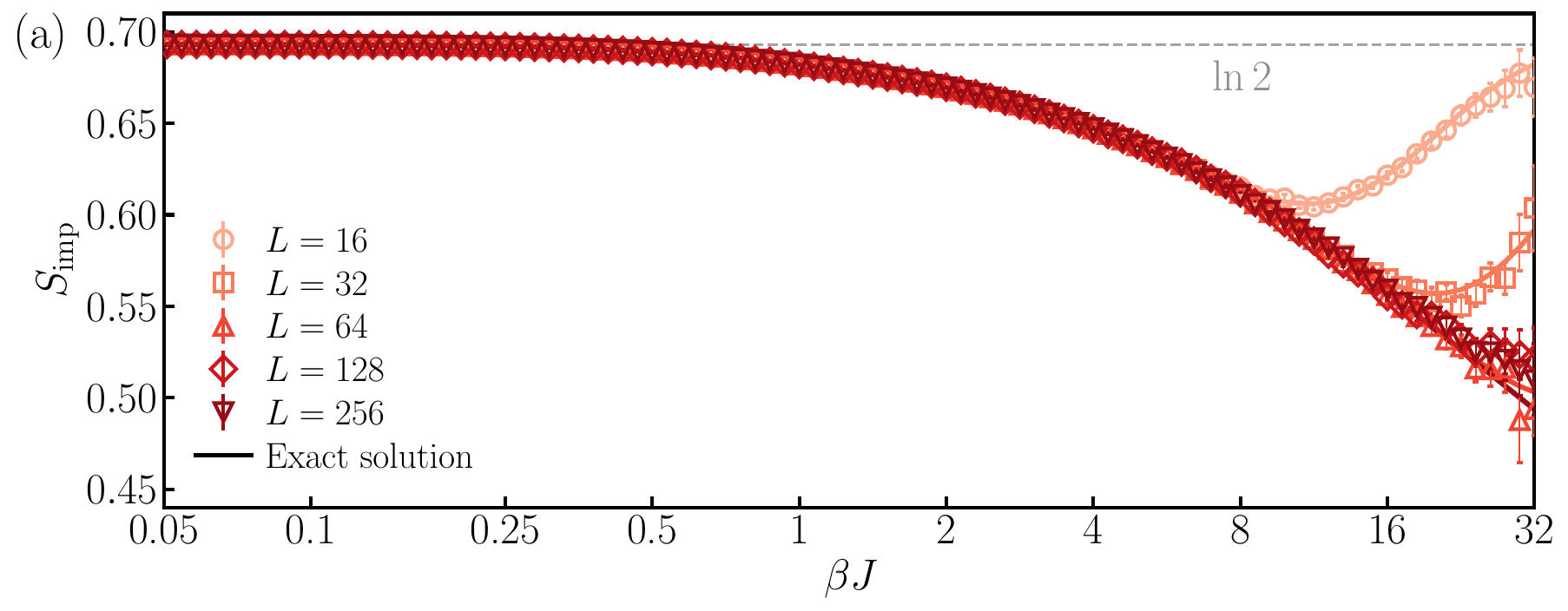}}
\hfill
\subfigure{\includegraphics[width=0.495\textwidth]{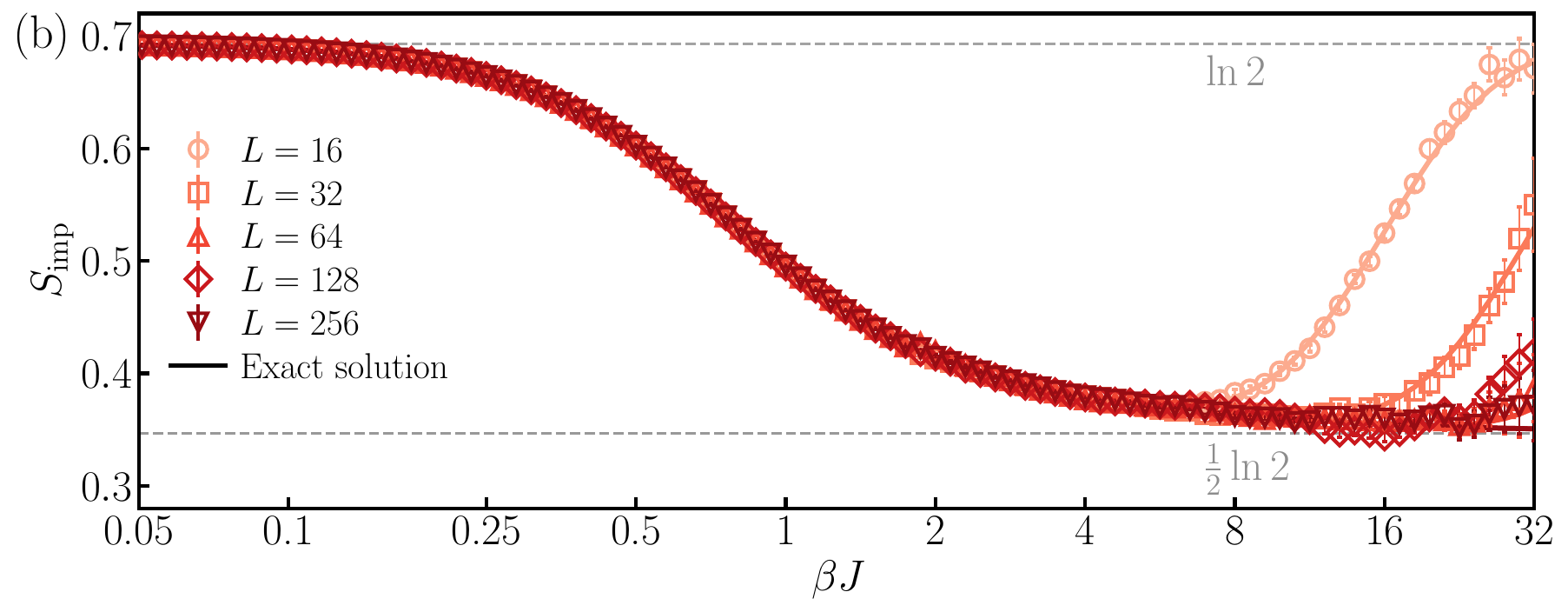}}
\caption{\justifying
Results for impurity entropy $S_{\text{imp}}$ of the TFIM as a function of inverse temperature $\beta$ for (a) $\lambda=0.2$ and (b) $\lambda=1.0$. 
The horizontal dashed lines mark $\ln 2$ and $\frac{1}{2}\ln 2$.
}
\label{fig_sm:TFIM}
\end{figure}

%
We perform additional high temperature Local-Maxwell QMC simulations for bath lengths $L=16,32,64,128,256$, extending the inverse-temperature range down to $\beta=0.05$.
The reconstructed impurity entropies for $\lambda=0.2$ and $1.0$ are compared with the corresponding exact finite-size solutions in Fig.~\ref{fig_sm:TFIM}.
%

\subsection{XXZ Model}
%

To quantify the statistical efficiency of the Local-Maxwell method, we define the cost-weighted variance as
\begin{equation}
\mathcal{V}=\mathcal{C}\sigma_S^2,
\end{equation}
where $\mathcal{C}$ denotes the total allocated CPU core time and $\sigma_S$ is the statistical standard error of the impurity entropy. 
For Monte Carlo estimates with $\sigma_S^2\propto1/\mathcal{C}$, a smaller $\mathcal{V}$ indicates higher statistical efficiency at a given computational cost. 
%
We estimate the computational cost from the allocated CPU core time 
\begin{equation}
\mathcal{C}=N_{\mathrm{core}}\sum_{k\in\mathcal{P}}t_k,
\end{equation}
%
where $t_k$ is the elapsed time of the slowest MPI rank in the $k$-th task.
All simulations used $N_{\mathrm{core}}=24$.

To characterize the system-size dependence, we fit the cost-weighted variance $\mathcal{V}$ to the power-law form
\begin{equation}
\mathcal{V}=AL^p
\end{equation}
over $16\leq L\leq1024$.
The fitted exponents are summarized in Table~\ref{tab:XXZ_scaling}, with coefficients of determination close to $0.99$ for all cases.
The conventional extensive TI method exhibits an approximately quadratic growth, whereas the Local-Maxwell method shows a much weaker, nearly linear size dependence.
%
\begin{table}[htb]
\centering
\caption{Power-law exponents for the cost-weighted variance $\mathcal{V}=AL^p$ in the XXZ model over $16\leq L\leq1024$.}
\label{tab:XXZ_scaling}
\renewcommand{\arraystretch}{1.3}
\setlength{\tabcolsep}{12pt}
\begin{tabular}{cccc}
\hline\hline
$C_x$ & $\beta$ & $p_{\mathrm{LM}}$ & $p_{\mathrm{TI}}$\\
\hline
$0.2$ & $8$  & $1.069$ & $2.029$\\
$0.2$ & $32$ & $1.313$ & $2.021$\\
$1.0$ & $8$  & $1.018$ & $2.063$\\
$1.0$ & $32$ & $1.192$ & $1.969$\\
\hline\hline
\end{tabular}
\end{table}

\begin{figure}[htbp]
\centering
\subfigure{\includegraphics[width=0.325\textwidth]{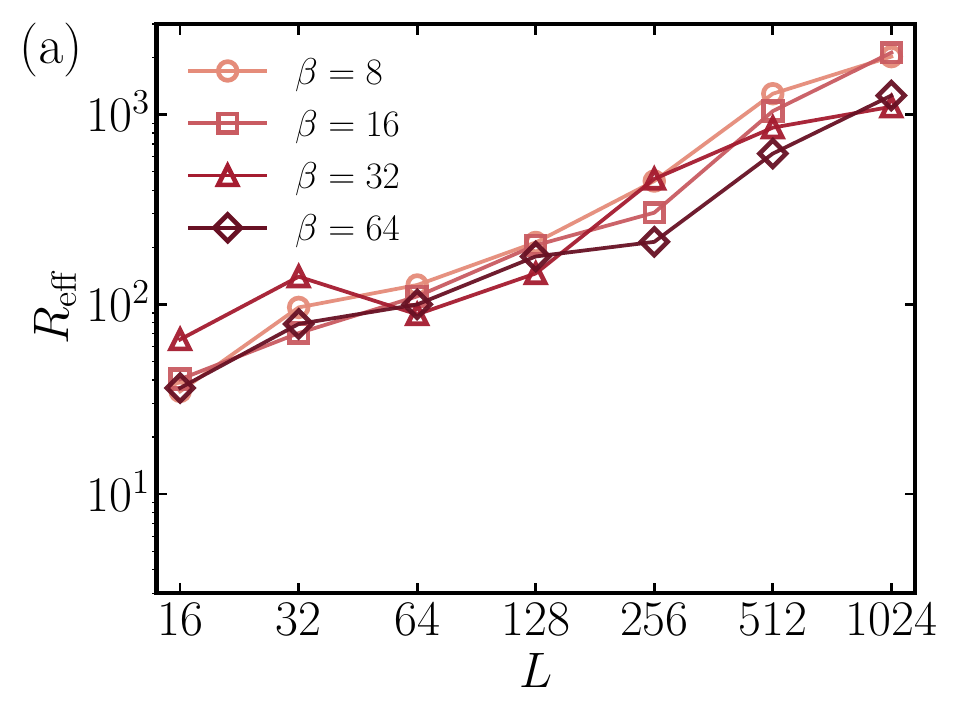}}
\subfigure{\includegraphics[width=0.325\textwidth]{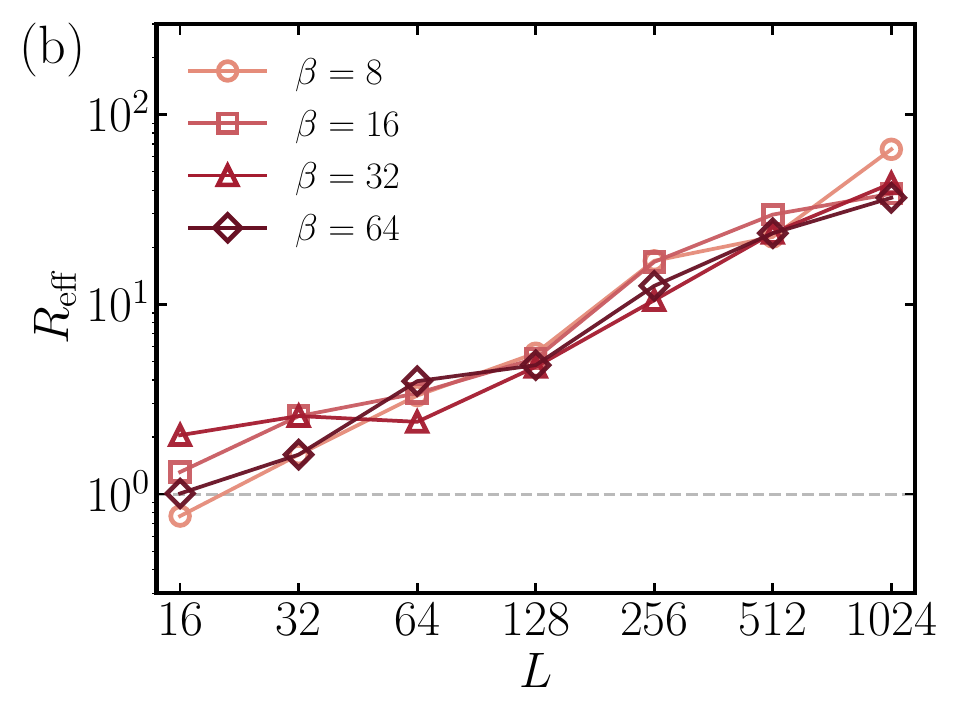}}
\caption{\justifying
Relative statistical efficiency $R_{\text{eff}}$ of the two methods versus system size $L$ 
for the XXZ spin chain, (a) $C_x=0.2$ and (b) $C_x=1.0$.
The horizontal dashed line marks $R_{\mathrm{eff}}=1$.
}
\label{fig_sm:XXZ_efficiency}
\end{figure}

To directly compare the statistical efficiency of the Local-Maxwell method and extensive thermodynamic integration (TI), we define the relative efficiency as
\begin{equation}
  R_{\text{eff}}=\frac{\mathcal{V}_{\text{TI}}}{\mathcal{V}_{\text{LM}}},
\end{equation}
where $R_{\text{eff}}>1$ indicates a statistical advantage of the Local-Maxwell method. 
The fitted exponents imply $R_{\text{eff}}\propto L^{\Delta p}$, with $\Delta p=p_{\mathrm{TI}}-p_{\mathrm{LM}}>0$ for all cases considered. 
As shown in Fig.~\ref{fig_sm:XXZ_efficiency}, the relative efficiency increases with system size for both impurity couplings, 
consistent with the more favorable scaling of the Local-Maxwell estimator.

\subsection{Three-channel XXZ impurity model}
In this section, we analyze the weak and strong coupling limits of the three-channel XXZ impurity model 
and present a free energy based consistency check of the entropy recovery. 
The Hamiltonian is
\begin{equation}
  \mathcal{H}=J\sum_{\alpha=1}^{3}\sum_{j=1}^{L-1}\left(S_{\alpha,j}^{x}S_{\alpha,j+1}^{x}+S_{\alpha,j}^{y}S_{\alpha,j+1}^{y}+\Delta S_{\alpha,j}^{z}S_{\alpha,j+1}^{z}\right)
  +J_K\sum_{\alpha=1}^{3}\left(S_d^xS_{\alpha,1}^x+S_d^yS_{\alpha,1}^y+\Delta S_d^zS_{\alpha,1}^z\right),
  \label{eq:sm_HM3xxz}
\end{equation}
where we set $J=1$ and $\Delta=0.8$, while $J_K$ controls the impurity-bath coupling.

\subsubsection{High-temperature expansion}
We first consider the high-temperature limit $\beta J\ll1$. 
By writing $\mathcal H=\mathcal H_0+J_KV$, where $V=\partial_{J_K}\mathcal H$ denotes the impurity exchange operator, 
we expand the impurity partition-function ratio around $\beta=0$.
At infinite temperature, $\langle V\rangle_0=0$, and the second moment is
\begin{equation}
\langle V^2\rangle_0=\sum_{\alpha=1}^{3}\left\langle\left(S_d^xS_{\alpha,1}^x+S_d^yS_{\alpha,1}^y+\Delta S_d^zS_{\alpha,1}^z\right)^2\right\rangle_0=\frac{3}{16}(2+\Delta^2),
\end{equation}
where cross terms between different chains vanish under the infinite-temperature trace.
The cumulant expansion gives
\begin{equation}
\ln\frac{\mathcal{Z}_{\mathrm{junction}}}{\mathcal{Z}_{\mathrm{bath}}}=\ln2+\frac{\beta^2J_K^2}{2}\langle V^2\rangle_0+\mathcal O(\beta^3).
\end{equation}
Using $S=(1-\beta\partial_\beta)\ln \mathcal{Z}$, we obtain
\begin{equation}
S_{\mathrm{imp}}=\ln2-\frac{3}{32}(2+\Delta^2)(\beta J_K)^2+\mathcal O(\beta^3).
\label{eq:sm_M3_highT}
\end{equation}

For $\Delta=0.8$, the leading coefficient is $3(2+\Delta^2)/32=0.2475$.
The initial entropy depletion is therefore quadratic in $\beta J_K$, with a coefficient determined entirely by the local impurity exchange and independent of the bath size.

\begin{figure}[htbp]
\centering
\subfigure{\includegraphics[width=0.325\textwidth]{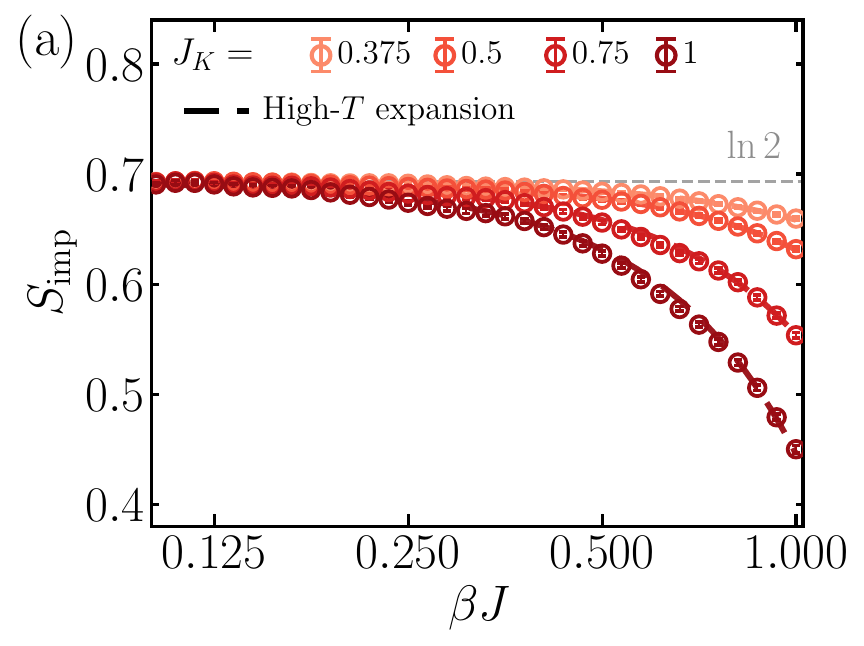}}
\subfigure{\includegraphics[width=0.325\textwidth]{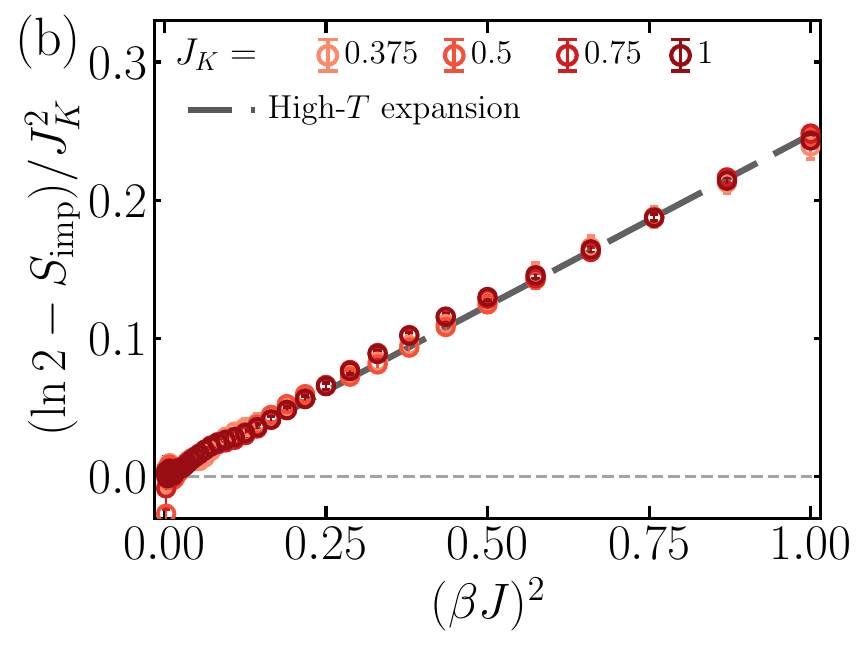}}
\caption{\justifying
High-temperature impurity entropy of the three-channel XXZ model at $\Delta=0.8$ and $L=512$.
(a) Impurity entropy $S_{\mathrm{imp}}$ as a function of $\beta J$ for $J_K=0.375,0.5,0.75$, and $1.0$.
The horizontal dashed line marks $\ln 2$.
(b) Rescaled entropy depletion $(\ln2-S_{\mathrm{imp}})/J_K^2$ versus $(\beta J)^2$.
Dashed lines represent the leading order high temperature expansion in Eq.~\eqref{eq:sm_M3_highT}.
}
\label{fig_sm:M3_highT}
\end{figure}

Fig.~\ref{fig_sm:M3_highT} compares this prediction with the Local-Maxwell results for $L=512$ and different impurity couplings over $0.05\lesssim\beta J\leq1$.
The numerical results exhibit the expected initial entropy depletion and an approximate collapse when $(\ln2-S_{\mathrm{imp}})/J_K^2$ is plotted against $(\beta J)^2$, consistent with Eq.~\eqref{eq:sm_M3_highT}.
The dashed lines show the leading-order expansion extended over the displayed temperature range for comparison; quantitative agreement is expected only in the high-temperature regime.

\subsubsection{Weak-coupling limit}
When $J_K\ll J$, the three XXZ chains are described by independent Luttinger liquids.
For the spin-$1/2$ XXZ chain at zero magnetic field and $-1<\Delta<1$, the Luttinger parameter is
\begin{equation}
K=\frac{\pi}{2(\pi-\arccos\Delta)}.
\end{equation}
This gives $K\simeq0.629$ at $\Delta=0.8$. 
The transverse boundary spin operator has scaling dimension $d_\perp=1/2K\simeq0.795$.

The transverse impurity exchange is therefore relevant at the decoupled fixed point, with
\begin{equation}
\frac{\mathrm{d}g_\perp}{\mathrm{d}\ell}
=(1-d_\perp)g_\perp+\mathcal O(g_\perp g_z,g_\perp^3),
\end{equation}
where $g_\perp$ and $g_z$ are dimensionless transverse and longitudinal couplings.
The longitudinal exchange has boundary scaling dimension one and is marginal at tree level.

To examine the corresponding entropy response, we expand the impurity free energy around $J_K=0$.
Writing $V=\partial_{J_K}\mathcal H$, the first-order term vanishes because the decoupled system satisfies $\langle V\rangle_{J_K=0}=0$.
The second-order cumulant expansion gives
\begin{equation}
F_{\mathrm{imp}}(T,J_K)
=-T\ln2-\frac{J_K^2}{2}\chi_V(T)+\mathcal O(J_K^3),
\end{equation}
where the susceptibility is
\begin{equation}
\chi_V(T)=\int_0^\beta\mathrm{d}\tau\,\langle V(\tau)V(0)\rangle_0.
\end{equation}
Writing $V=V_\perp+\Delta V_z$, the mixed transverse-longitudinal correlation vanishes by $U(1)$ symmetry, giving
\begin{equation}
\chi_V(T)=\chi_\perp(T)+\Delta^2\chi_z(T),
\end{equation}
where $\chi_\perp$ and $\chi_z$ denote the corresponding integrated correlation functions.
The longitudinal contribution is subleading to the transverse singular term.
The impurity entropy follows as
\begin{equation}
S_{\mathrm{imp}}(T,J_K)
=\ln2+\frac{J_K^2}{2}\partial_T\chi_V(T)+\mathcal O(J_K^3).
\end{equation}

At the decoupled fixed point, the transverse boundary correlation decays as
\begin{equation}
\langle S_{\alpha,1}^+(\tau)S_{\alpha,1}^-(0)\rangle_0
\sim|\tau|^{-2d_\perp}.
\end{equation}
For $1/2<d_\perp<1$, its singular contribution to the integrated susceptibility takes the form
\begin{equation}
\chi_\perp(T)=\chi_\perp(0)-BT^{2d_\perp-1}+\cdots,
\qquad B>0.
\end{equation}
The leading singular entropy correction is therefore
\begin{equation}
S_{\mathrm{imp}}-\ln2
\sim-AJ_K^2T^{2d_\perp-2},
\qquad A>0.
\label{eq:sm_weak_entropy}
\end{equation}
For $\Delta=0.8$, the exponent is $2d_\perp-2\simeq-0.410$.
The negative correction describes enhanced impurity-entropy depletion upon cooling.
This perturbative form applies in the weak coupling regime above the crossover scale at which the renormalized transverse coupling becomes strong.

\subsubsection{Strong-coupling limit}
When $J_K\gg J$, the impurity spin and the three adjacent endpoint spins form a four-spin cluster described by
\begin{equation}
\mathcal H_{\mathrm{cl}}=J_K\sum_{\alpha=1}^{3}\left(S_d^xS_{\alpha,1}^x+S_d^yS_{\alpha,1}^y+\Delta S_d^zS_{\alpha,1}^z\right).
\end{equation}
By introducing the total endpoint spin $\mathbf T=\sum_{\alpha=1}^{3}\mathbf S_{\alpha,1}$, the cluster Hamiltonian becomes [see Eq.~\eqref{eq:M3XXZ_T_sm}]
\begin{equation}
\mathcal H_{\mathrm{cl}}=J_K\left[\frac12(S_d^+T^-+S_d^-T^+)+\Delta S_d^zT^z\right].
\end{equation}
Since $\mathcal H_{\mathrm{cl}}$ commutes with $\mathbf T^2$ and $M=S_d^z+T^z$, its spectrum separates into $T=1/2$ and $T=3/2$ sectors.
For $\Delta=0.8$, the unique ground state belongs to the $T=3/2$, $M=0$ sector, with energy and corresponding state
\begin{equation}
E_0=-J_K\left(1+\frac{\Delta}{4}\right)=-1.2J_K, \qquad 
|0\rangle=\frac{1}{\sqrt2}\left(\left|\frac12,-\frac12\right\rangle-\left|-\frac12,\frac12\right\rangle\right)
\end{equation}
where the entries denote $S_d^z$ and $T^z$, respectively.
This state coincides with the $S=1$, $M=0$ state of the isotropic cluster and is not a total spin-singlet.
The lowest excited states belong to the $T=3/2$, $M=\pm1$ sectors, with energy
\begin{equation}
E_{\pm1}=-J_K\left(\frac{\Delta}{4}+\frac{\sqrt{\Delta^2+3}}{2}\right).
\end{equation}
The excitation gap of the cluster $\mathcal H$ is therefore
\begin{equation}
\delta_{\mathrm{cl}}=E_{\pm1}-E_0=J_K\left(1-\frac{\sqrt{\Delta^2+3}}{2}\right)\simeq0.04606J_K.
\label{eq:sm_cluster_gap}
\end{equation}

The cluster is coupled to the remaining chains through
\begin{equation}
\mathcal H_{\mathrm{link}}=J\sum_{\alpha=1}^{3}\left(S_{\alpha,1}^xS_{\alpha,2}^x+S_{\alpha,1}^yS_{\alpha,2}^y+\Delta S_{\alpha,1}^zS_{\alpha,2}^z\right).
\end{equation}
For $J\ll\delta_{\mathrm{cl}}$, we project onto the cluster ground state using $\mathcal P=|0\rangle\langle0|$ and $\mathcal Q=1-\mathcal P$.
The first-order contribution vanishes since $\langle0|\mathbf S_{\alpha,1}|0\rangle=0$.
The second-order Schrieffer-Wolff contribution is
\begin{equation}
\mathcal H_{\mathrm{eff}}^{(2)}=-\mathcal P\mathcal H_{\mathrm{link}}\mathcal Q\frac{1}{\mathcal Q\mathcal H_{\mathrm{cl}}\mathcal Q-E_0}\mathcal Q\mathcal H_{\mathrm{link}}\mathcal P.
\end{equation}
Up to an additive constant, the induced interactions between the remaining chain endpoints take the form
\begin{equation}
\mathcal H_{\mathrm{inter}}^{\mathrm{eff}}=\sum_{\alpha<\beta}\left[J_{xy}^{\mathrm{eff}}\left(S_{\alpha,2}^xS_{\beta,2}^x+S_{\alpha,2}^yS_{\beta,2}^y\right)+J_z^{\mathrm{eff}}S_{\alpha,2}^zS_{\beta,2}^z\right].
\end{equation}
For $\alpha\ne\beta$, the effective couplings are determined by virtual cluster excitations,
\begin{align}
J_{xy}^{\mathrm{eff}}&=-2J^2\sum_{n\ne0}\frac{\langle0|S_{\alpha,1}^x|n\rangle\langle n|S_{\beta,1}^x|0\rangle}{E_n-E_0},\\
J_z^{\mathrm{eff}}&=-2J^2\Delta^2\sum_{n\ne0}\frac{\langle0|S_{\alpha,1}^z|n\rangle\langle n|S_{\beta,1}^z|0\rangle}{E_n-E_0}.
\end{align}
At $\Delta=0.8$, these expressions yield
\begin{equation}
J_{xy}^{\mathrm{eff}}=-\frac{325}{42}\frac{J^2}{J_K},\qquad
J_z^{\mathrm{eff}}=\frac4{15}\frac{J^2}{J_K}.
\end{equation}
At the decoupled open-chain fixed point, the transverse interchain exchange has scaling dimension $1/K>1$, while the longitudinal exchange has dimension two.
Both perturbations are therefore irrelevant for $\Delta=0.8$.
Consequently, the asymptotic strong coupling fixed point consists of three effectively decoupled XXZ chains of length $L-1$.

Using the low-temperature bulk entropy density $s_{\mathrm{bulk}}(T)=\pi T/(3v)$ of a critical XXZ chain, we obtain
\begin{equation}
S_{\mathrm{imp}}(T)\simeq3\left[S_{\mathrm{chain}}(L-1,T)-S_{\mathrm{chain}}(L,T)\right]\simeq-\frac{\pi T}{v},
\label{eq:sm_strong_entropy}
\end{equation}
where $v=\pi J\sqrt{1-\Delta^2}/(2\arccos\Delta)$ is the spin velocity. 
This asymptotic expression applies for $v/L\ll T\ll\min(J,\delta_{\mathrm{cl}})$, provided that $J\ll\delta_{\mathrm{cl}}$. 

The excess entropy is thus negative and decreases in magnitude upon cooling.
Since our simulations cover $J_K\leq J$, this strong coupling analysis serves as an asymptotic reference. 

%
\subsubsection{Free energy chord reconstruction}
The free energy chord analysis in the main text is based on impurity free energies reconstructed from local observables, avoiding numerical temperature differentiation.
For $\mathcal H=\mathcal H_0+J_KV$, the local conjugate observable is
\begin{equation}
\langle V\rangle=-\frac{\langle N_D\rangle}{\beta J_K}+\Delta\langle A_D\rangle,
\end{equation}
where $A_D$ is the imaginary-time averaged longitudinal impurity exchange summed over the three chains.
The impurity free energy follows from
\begin{equation}
F_{\mathrm{imp}}(T,J_K)=-T\ln2+\int_0^{J_K}\mathrm{d}j\,\langle V\rangle_{T,j}.
\end{equation}
The coupling integration uses the same quadrature as in the Local-Maxwell reconstruction.
The chord entropy is obtained from the free energy difference
\begin{equation}
S_{\mathrm{ch}}(T_1,T_2)=-\frac{F_{\mathrm{imp}}(T_2,J_K)-F_{\mathrm{imp}}(T_1,J_K)}{T_2-T_1}.
\end{equation}
For $J=1$, we denote $F_b=F_{\mathrm{imp}}(1/b,J_K)$ and obtain the cold, middle, and warm chord entropies as
\begin{align}
S_{\mathrm c}&=32(F_{32}-F_{16}),\\
S_{\mathrm m}&=16(F_{16}-F_8),\\
S_{\mathrm w}&=8(F_8-F_4).
\end{align}
The adjacent-window contrasts are $D_{\mathrm{low}}=S_{\mathrm c}-S_{\mathrm m}$ and $D_{\mathrm{high}}=S_{\mathrm m}-S_{\mathrm w}$, 
satisfying $D_{\mathrm{ch}}=D_{\mathrm{low}}+D_{\mathrm{high}}=S_{\mathrm c}-S_{\mathrm w}$.
These quantities are constructed directly from free energy differences without numerical temperature derivatives.

\begin{figure}[htbp]
\centering
\includegraphics[width=0.70\textwidth]{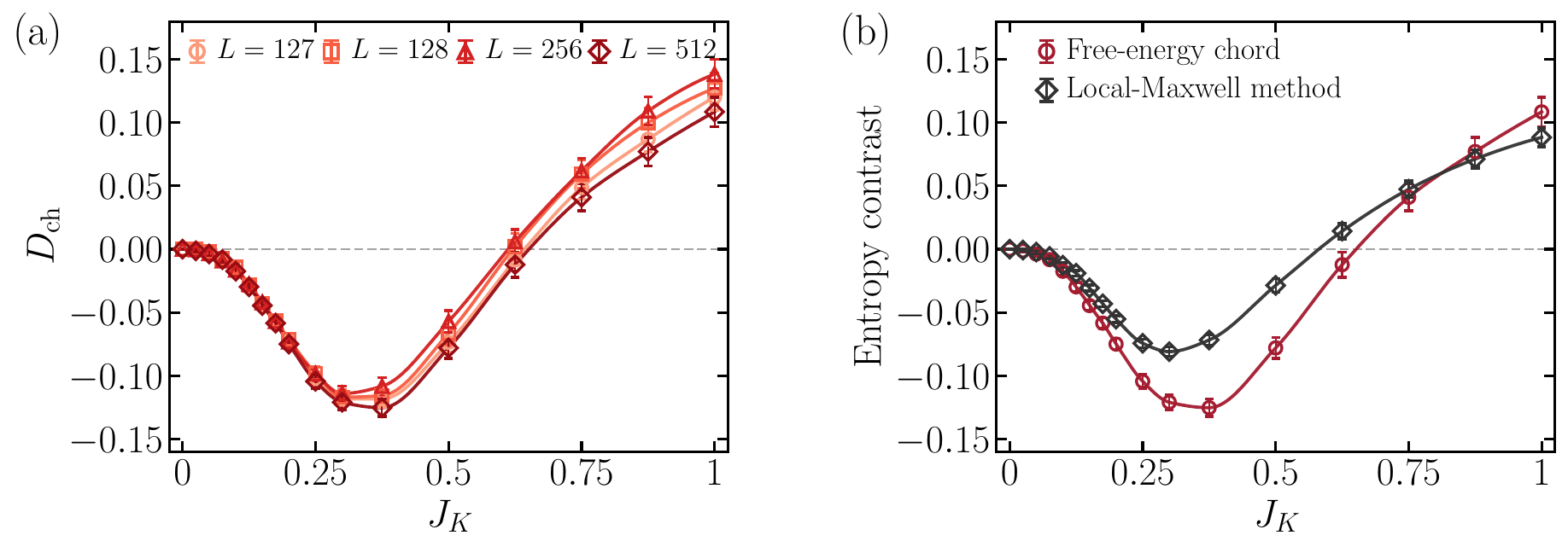}
\caption{\justifying
Additional consistency checks for the three-channel XXZ impurity model.
(a) Free energy chord contrast $D_{\mathrm{ch}}$ for $L=127,128,256,512$.
(b) Comparison between $D_{\mathrm{ch}}$ and the Local-Maxwell entropy contrast $D_{\mathrm{LM}}$ for $L=512$, using their respective temperature averages.
}
\label{fig_sm:M3_chord}
\end{figure}

As an additional consistency check, Fig.~\ref{fig_sm:M3_chord}(a) presents the chord contrast for different system sizes, including neighboring odd and even chain lengths.
Fig.~\ref{fig_sm:M3_chord}(b) compares $D_{\mathrm{ch}}$ with the Local-Maxwell contrast $D_{\mathrm{LM}}=\overline{S}_{\mathrm{imp}}^{\mathrm{cold}}-\overline{S}_{\mathrm{imp}}^{\mathrm{warm}}$ at $L=512$.
Although these two contrasts employ different temperature averages, both exhibit a sign reversal with increasing $J_K$, supporting the finite-temperature entropy recovery.

\subsection{Details of the Worldline QMC}
We provide an exact diagonalization (ED) benchmark for the TFIM and summarize the numerical reconstruction procedures 
used in our worldline QMC calculations.

\subsubsection{Benchmark of Local-Maxwell QMC method}
\begin{figure}[htbp]
\centering
\subfigure{\includegraphics[width=0.325\textwidth]{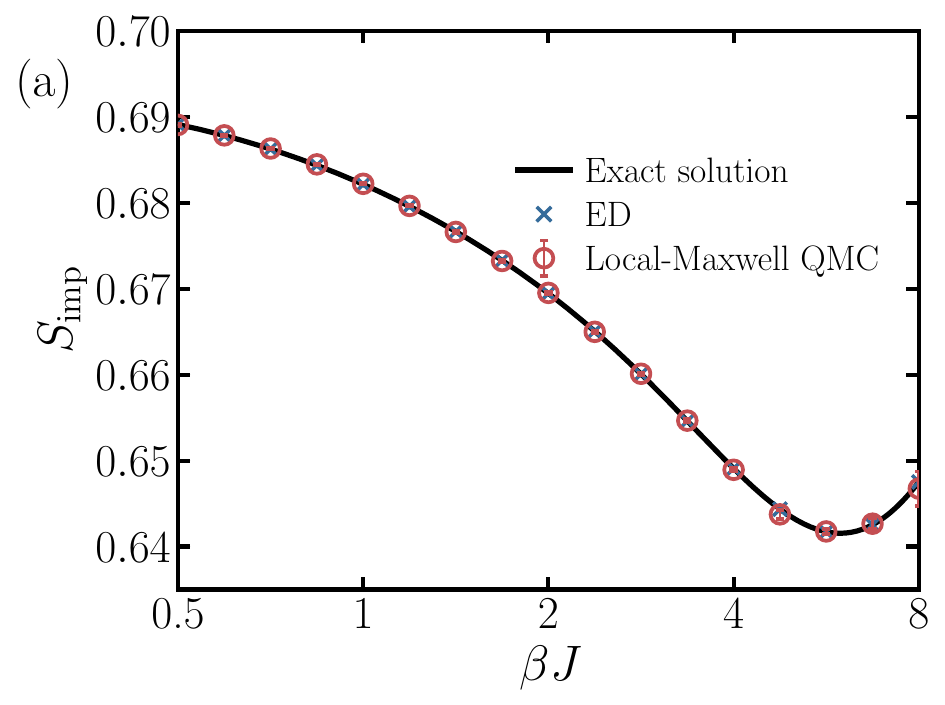}}
\subfigure{\includegraphics[width=0.325\textwidth]{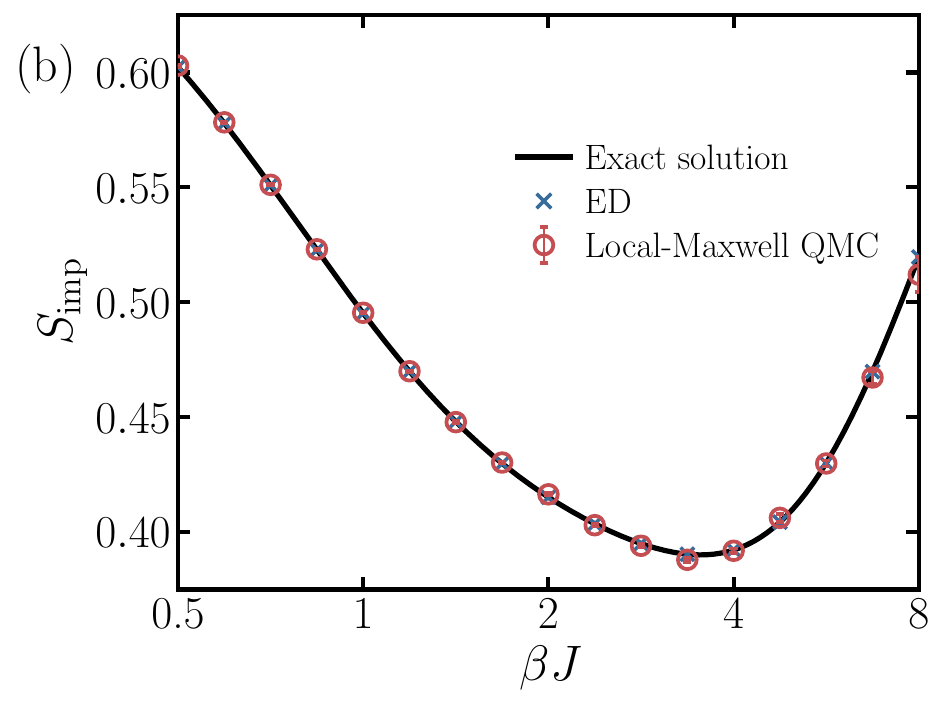}}
\caption{\justifying
Benchmarks for impurity entropy $S_{\text{imp}}$ of the Local-Maxwell method for the critical TFIM with $L=8$ bath spins. 
(a) and (b) are the cases of $\lambda = 0.2$ and $1.0$, respectively. 
}
\label{fig_sm:TFIM_ED}
\end{figure}

We benchmark the Local-Maxwell method against independent ED of the critical TFIM with $L=8$ bath spins. 
As shown in Fig.~\ref{fig_sm:TFIM_ED}, the reconstructed impurity entropy agrees with the exact results within statistical uncertainties for $\lambda=0.2$ and $1.0$.
%

\subsubsection{Numerical reconstruction of impurity entropy}
We employ worldline QMC to sample the local observables entering the Maxwell relation. For $\mathcal H(g)=\mathcal H_0+gV$, the impurity entropy is reconstructed as
\begin{equation}
S_{\mathrm{imp}}(\beta,g)=\ln 2+\int_0^g \mathrm{d}g'\,(1-\beta\partial_\beta)Q_{g'}(\beta),
\end{equation}
where $Q_g=\partial_g\ln\mathcal Z=-\beta\langle V\rangle$, and the temperature derivative is taken at fixed coupling.
The initial entropy $\ln2$ corresponds to the decoupled spin-$1/2$ impurity. 

For the TFIM and three-channel XXZ models, we use logarithmically spaced $\beta$-grids, with uniform spacing in $\ln\beta$, 
and evaluate the temperature derivatives using local polynomial fits of up to third and fifth order, respectively.
For the XXZ chain, the temperature derivative is instead evaluated using three neighboring points, $\beta e^{-0.1}$, $\beta$, and $\beta e^{0.1}$, through quadratic interpolation in $\ln\beta$.
Besides, for the three-channel XXZ model, the coupling integration is performed before numerical temperature differentiation.

Statistical uncertainties are estimated from independent Monte Carlo measurements and propagated through the numerical temperature differentiation and coupling integration.
For the free energy chord analysis, uncertainties are propagated from the corresponding free energy estimates.

\end{document}